\documentclass[showpacs,preprintnumbers,10pt,twocolumn]{revtex4-1}
\usepackage{dcolumn}
\usepackage{bm}
\usepackage{graphicx}
\usepackage{epstopdf}

\begin{document}

\title{Optical hyperpolarization of $^{13}C$ nuclear spins in nanodiamond ensembles}

\author{Q. Chen$^{1,2}$, I. Schwarz$^{1,2}$, F. Jelezko$^{2,3}$, A. Retzker$^{4}$ and M.B. Plenio$^{1,2}$ }

\affiliation{$^{1}$ Institut f\"{u}r Theoretische Physik, Albert-Einstein-Allee 11, Universit\"{a}t Ulm, 89069 Ulm, Germany\\ $^{2}$ IQST,  Albert-Einstein-Allee 11, Universit\"{a}t Ulm, 89069 Ulm, Germany\\ $^{3}$ Institut f\"{u}r Quantenoptik, Universit{\"a}t Ulm, 89073 Ulm, Germany\\
$^{4}$ Racah Institute of Physics, The Hebrew University of Jerusalem, Jerusalem, 91904, Israel}

\begin{abstract}
%
Dynamical nuclear polarization holds the key for orders of magnitude enhancements
of nuclear magnetic resonance signals which, in turn, would enable a wide range of
novel applications in biomedical sciences. However, current implementations of DNP
require cryogenic temperatures and long times for achieving high polarization.
Here we propose and analyse in detail protocols that can achieve rapid hyperpolarization
of $^{13}C$ nuclear spins in randomly oriented ensembles of nanodiamonds at room
temperature. Our protocols exploit a combination of optical polarization of electron
spins in nitrogen-vacancy centers and the transfer of this polarization to $^{13}C$
nuclei by means of microwave control to overcome the severe challenges that are posed
by the random orientation of the nanodiamonds and their nitrogen-vacancy centers.
Specifically, these random orientations result in exceedingly large energy variations of
the electron spin levels that render the polarization and coherent control of the nitrogen-vacancy
center electron spins as well as the control of their coherent interaction with the surrounding
$^{13}C$ nuclear spins highly inefficient. We address these challenges by a combination of an
off-resonant microwave double resonance scheme in conjunction with a realisation of the integrated
solid effect which, together with adiabatic rotations of external magnetic fields or rotations of
nanodiamonds, leads to a protocol that achieves high levels of hyperpolarization of the entire
nuclear-spin bath in a randomly oriented ensemble of nanodiamonds even at room temperature. This
hyperpolarization together with the long nuclear spin polarization lifetimes in nanodiamonds and
the relatively high density of $^{13}C$ nuclei has the potential to result in a major signal enhancement
in $^{13}C$ nuclear magnetic resonance imaging and suggests functionalized and hyperpolarized
nanodiamonds as a unique probe for molecular imaging both in vitro and in vivo.

\end{abstract}

\maketitle
\section{Introduction}
Nuclear magnetic resonance (NMR)~\cite{Bloch,Bloch2} and magnetic resonance imaging (MRI) \cite{Lauterbur}
have evolved to be powerful techniques to extract molecular-level information in a wide variety of
physical, chemical and biological applications. Recently, $^{13}C$ based MRI has emerged as a new platform
enabling essentially background-free imaging of non-proton nuclei with the additional possibility for the
local and permanent destruction of the signal by means of radio frequency (RF) pulses~\cite{Review}. However,
the low signal-to-noise ratio for $^{13}C$ based MRI is not sufficient for most clinical and research
applications due to the combination of a relatively low gyromagnetic ratio of $^{13}C$ and of the low
natural abundance of this nucleus. Several strategies have been proposed to enhance the sensitivity of
$^{13}C$ based MRI. Particularly promising in this context is the hyperpolarization of the $^{13}C$ nuclei,
i.e. the generating of a large, non-thermal $^{13}C$ nuclear spin-polarization. Indeed, hyperpolarized
$^{13}C$ based MRI has shown exciting potential for in vivo applications, especially for novel metabolic
imaging~\cite{Dutta2,Gallagher1,Gallagher2}. One of the most powerful methods for the generation of hyperpolarization
is dynamic nuclear polarization (DNP)~\cite{Fridlund1,Fridlund3,Fridlund4,Fridlund5,Fridlund6}, in which
a large polarization of electron spins is transferred to nuclear spins, resulting in an enhancement of the
MRI signal by several orders of magnitude.

Over the last decade, several breakthroughs have occurred in the manufacturing, surface treatment and
use of nanoparticles for biomedical applications~\cite{Pankhurst}. Biocompatible nanoparticles present
an attractive platform for hyperpolarization, as they can be functionalized for molecular specificity,
ca exhibit long nuclear spin relaxation times, and contain numerous nuclear spins of a single species (e.g.
carbon). Long-lived hyperpolarization has been demonstrated in silica nanopartices~\cite{Aptekar} and
nanodiamonds~\cite{Dutta,Rejy}, using standard DNP protocols, i.e. transferring the high thermal electron
spin polarization at cryogenic temperatures in a strong external magnetic field to nuclei. Nanodiamonds
in particular offer exciting possibilities as novel hyperpolarized probes. In addition to excellent
biocompatibility~\cite{Mochalin}, the possibility for surface functionalisation and nuclear spin relaxation
time of several minutes~\cite{Casabianca}, nanodiamonds contain crystal point-defects with unique optical
and magnetic properties. Amongst these the negatively charged nitrogen-vacancy (NV) center stands out as
its electron spin can be polarized within microseconds by optical pumping while exhibiting a relaxation
time in the millisecond range even at room temperature~\cite{Jelezko}. This establishes the NV center
electron spin as promising candidate for achieving nanodiamond hyperpolarization at ambient conditions
- room-temperature optical DNP.

Previous work has demonstrated that in bulk diamond electron spin polarization of the NV center can be
generated and subsequently be transferred via hyperfine interactions to nearby nuclei. This polarization
can then be detected indirectly via the NV center~\cite{Dutta,Gurude,Jacques,Neumann,King,London,Togan,Wang,Bretschneider}
or directly via an NMR scanner~\cite{Fischer}. To this end, in bulk diamonds a high magnetic field is
aligned with the quantisation axis of the NV center to enable optical pumping to a specific electron
spin state. Controlled interaction of the electron spin with specific nearby nuclear spins species can
then be achieved for example by the application of continuous wave microwave (MW) radiation applied to
the NV center electron spin~\cite{London}. When the Rabi frequency
of the microwave driving field matches the Larmor frequency of a specific nuclear spin species (achieving
a Hartmann-Hahn (H-H) resonance~\cite{Hartman}), flip-flops can occur between the microwave dressed states
of the electron spin of the NV center and the surrounding $^{13}C$ nuclear spins~\cite{Cai,Cai2,CaiRJP13}
as was recently demonstrated experimentally~\cite{London}.


However, medical MRI applications require large ensembles of nanodiamonds. In such ensembles (powder
or solution) the angle between the external magnetic field and the natural orientations of NV centers
and therefore their quantisation axis is randomly distributed. This brings about several challenges
for achieving high levels of hyperpolarization that need to be addressed by means of carefully designed
protocols. First, the energy levels of the NV spin are then distributed over a large energy range of
possible values so that the applied microwave field will be resonantly coupled to an exceedingly small
fraction of the NV spins. Secondly, in an external magnetic field, additional limitations concern the
optical polarization of the NV spins. In particular, the NV spins will be initialized to different
states relative to the laboratory frame depending on their angle with the external magnetic field,
thus resulting in a small net polarization.

In this work we will show to address these challenges to achieve a radical enhancement of the
hyperpolarization in a nanodiamond ensemble. To this end we use an off-resonant driving and
the integrated solid effect (ISE) for the NV center spin, resulting in a robust spin polarization
transfer between the NV and nuclear spins. A very large fraction of the NV electron spins can be
coherently coupled to neighboring nuclear spins. Additionally, adiabatic rotation of the magnetic
fields, or the Brownian rotation of the nanodiamonds themselves (in powder or solution respectively),
is proposed for extending the polarization scheme to almost the entire nanodiamond ensemble. We
also discuss the effect of spin diffusion induced by the nuclear dipole-diploe interactions which
will further support the polarization of large volumes.

The paper is organized as follows. In Sec. \ref{Section II} we present the coherent coupling and
polarization of a single nuclear spin near a randomly oriented NV spin. This serves the introduction
of the principal challenges in nanodiamond ensemble polarization, the discussion of the robust
initialization and construction of suitable dressed states for the NV center electron spin and the
realisation of near resonant coupling with the nuclear spin by using off-resonant driving and the
integrated solid effect. In Sec. \ref{Adiabatic rotations for more polarizations}, we introduce
mechanisms to extend our polarization protocols to the entire nanodiamond ensemble, both for
nanodiamond powder and for nanodiamonds in a solution. In Sec. \ref{Effect of nuclear dipolar interaction},
we consider our protocol for multiple $^{13}C$ nuclear spins and include the effect of dipolar
coupling among the nuclear spins and in particular the benefits of nuclear spin diffusion. In
Sec. \ref{A interesting range of the NV orientations}
we demonstrate the applicability of our polarization schemes for another very broad range of NV center
orientations. In Sec. \ref{Depolarizing effect} we take into account the effect of depolarization
processes on the efficiency of our polarization protocol. Finally, the discussion and conclusion
parts are given in Sec. \ref{Discussion} and Sec. \ref{Conclusions}, respectively.

\begin{figure}[htb]
\center
\includegraphics[width=3.4in]{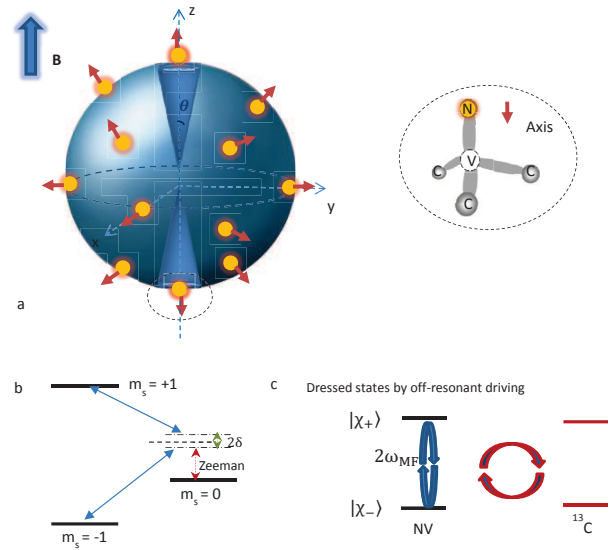}
\caption{Schematic of dressed-state resonant coupling of NV spins and nuclear spins in a nanodiamond
ensemble and energy-level diagrams. (a) The random orientations of the NV spins are uniformly distributed
over the unit sphere. Small yellow circles represent examples with orientations indicated by red arrows
which represent the unit vector pointing from the nitrogen to the vacancy of an NV center (see the
example of such an NV center in the dashed circle, the axis pointing along `N-V' forms the natural
quantization axis). (b) Ground electronic spin states of an NV center in a strong external magnetic
field. Microwave driving fields are applied to off-resonantly achieving an effective resonant double
quantum transition between the state $m_s=+1$ and $m_s=-1$ in the large detuning regime. (c) The
effective coupling provides a dressed state basis which then permits energy conserving flip-flops
between the dressed states and external resonant nuclear spins.}
\label{Schematic}
\end{figure}

\section{Magnetic manipulation and polarization of a single nuclear spin nearby a randomly oriented NV spin}
\label{Section II}

\subsection{Difficulties from the random orientations}
The nanodiamond ensembles that we are interested in are realised as powder or solutions. In both cases
the spatial orientations of the nanodiamonds and therefore of the NV centers hosted in them is random
and uniformly distributed across the full solid angle (see Fig. 1a for an illustration). In the laboratory
frame this leads to the lack of a common quantization axis for the NV centers. We begin by describing in
more detail the two principal challenges that are being imposed by these random spatial orientations:
(1) The direction of the natural quantization axis associated with the crystal-field energy splitting
$D$ is not controllable, resulting in a significant variance of the energy levels in the presence of
an external magnetic field and (2) optical pumping of the NV center electron spins in an external magnetic
field will not initialize all NV center electron spins to the same state.

\subsubsection{Zero-field and external magnetic field distribution}
\begin{figure}
\center
\includegraphics[width=2.8in]{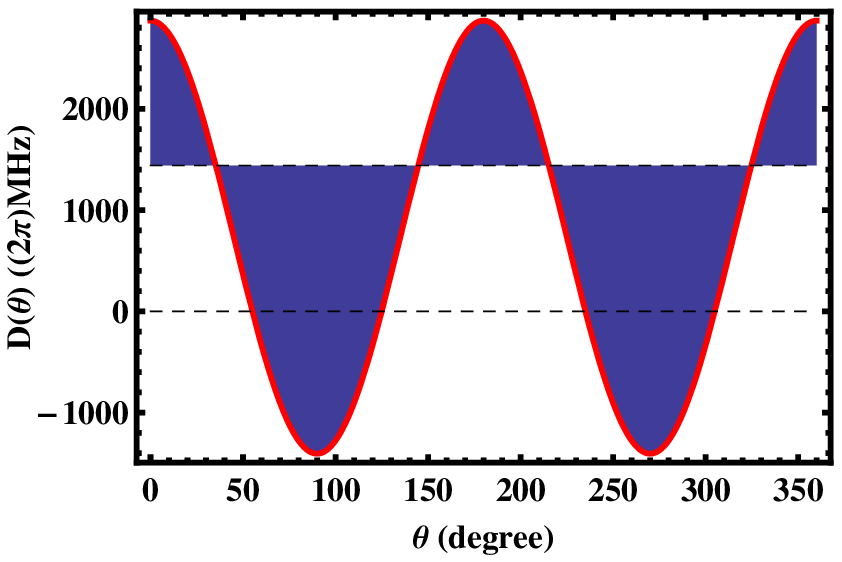}
\includegraphics[width=2.8in]{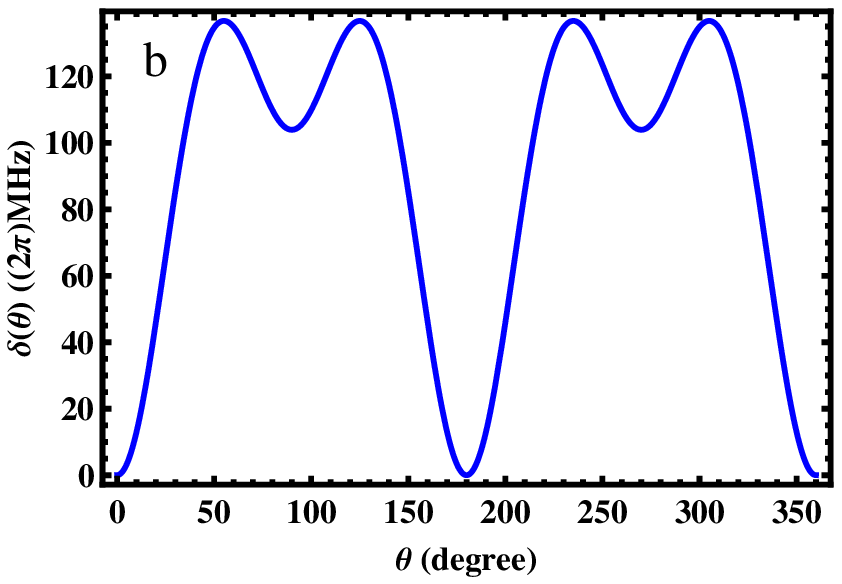}
\caption{a, Zero-field distribution $D(\theta)$ of the NV spins in nanodiamonds with
$D=(2\pi)2.87$ GHz and $E=(2\pi)20$ MHz. b, The second order corrections which induce a
energy distribution $\delta(\theta)$.}
\label{distribution}
\end{figure}

The negatively charged NV center, in the following denoted for brevity as the NV center, realises in its
ground state an electronic spin triplet ($S=1$) which exhibits a zero-field splitting of $D = (2\pi)2.87$
GHz which separates the $|m_s=0\rangle$ state energetically from the degenerate $|m_s=\pm\rangle$ manifold.
The application of an external magnetic field, in the following assumed to take the value $B=0.36$ T, lifts
this remaining degeneracy such that the state $m_s=-1$ is shifted below the $m_s=0$ state (see Fig.
\ref{Schematic}b). The zero-field splitting and the Zeemann effect due to an external magnetic field
of the NV center are described by
\begin{equation}
    H_{NV}=\vec{S}\mathbf{D}\vec{S} + \gamma_{e}\vec{B}\vec{S}.
\end{equation}
Herein $\mathbf{D}$ denotes the orientation dependent zero-field splitting tensor, $\gamma_e = (2\pi)
28.7\mbox{GHz}/\mbox{T}$ the gyromagnetic ratio, $\vec{S}$ the electron spin-1 vector operator~\cite{spin1}
and $\vec{B}$
the magnetic field vector. In the principal axis system defined by the NV symmetry axes, the zero-field
splitting tensor $\mathbf{D}$ is diagonal
\begin{equation}
    \mathbf{D}=diag(-\frac{1}{3}D+E,-\frac{1}{3}D-E,\frac{2}{3}D).
\end{equation}
Here $E$ denotes the strain dependent contribution. It is worth emphasizing that in the scenario that
we are considering here, the orientation of the symmetry axis of the NV center relative to the external
magnetic field is uniformly distributed over the unit sphere. For the following it will be convenient
to conduct the discussion in the laboratory frame whose z-axis we define to take the direction of the
externally applied strong magnetic field, $\gamma_{e}B\gg D$. In this frame, the zero-field splitting
tensor $\mathbf{D}$ will have off-diagonal elements. As $\gamma_{e}B\gg D$ these off-diagonal elements
are rapidly rotating so that their main effect will be energy shifts of the diagonal element of the
Hamiltonian. The Hamiltonian can then be written as (see the Appendix for details of the derivation)
\begin{equation} \label{hamNV}
    H''_{eff} = (\gamma_{e}B + \delta(\theta))S_{z} + D(\theta)S_{z}^{2}.
\end{equation}
where
\begin{eqnarray}
    D(\theta)&=&\frac{D(1+3\cos(2\theta))+3E(1-\cos(2\theta))}{4},\\ \nonumber
    \delta(\theta)&=&\frac{\gamma_e B|G_1|^2}{(\gamma_e B)^2-[D(\theta)]^2}+\frac{|G_2|^2}{2\gamma_e B},
\end{eqnarray}
$\theta$ is the angle between the magnetic field direction and the NV-axis and the $G_i$ are given in the
Appendix. Clearly, the random orientations of the NV centers cause a variation of the zero-field splitting
$D(\theta)$ across the entire interval $[-(2\pi)1.43\mbox{GHz},(2\pi)2.87 \mbox{GHz}]$ and $\delta(\theta)$
across the interval $[0 \mbox{MHz},(2\pi)140 \mbox{MHz}]$ as shown in Fig. \ref{distribution}.

If we were to follow the scheme of Ref.~\cite{London}, i.e. without the combination of off-resonant drive
and ISE technique that we will present in this work, the uncertain detuning of the MW frequency from the
electronic resonance can reach the order of GHz, which in turn would prevent effective polarization transfer.

\subsubsection{Optical initialization of NV spins at the high magnetic field limit}
A second important difference between a randomly oriented nanodiamond ensemble and a bulk
diamond concerns the optical polarization of electron spins of the NV center. For bulk
diamonds, the magnetic field can be aligned with the principal axis of the NV center and the
electronic spin of the NV center can be optically polarized to the state $|m_s=0\rangle$
by illumination with a $532$ nm green laser. However, for an ensemble of randomly oriented
nanodiamonds, even in the limit of a strong magnetic field, the NV centers will be optically
pumped to the state $|m_s=0\rangle_{\theta}$ that is defined by the relative orientation
of the NV center with respect to the externally applied magnetic field which defines the
laboratory frame. 

As discussed in Appendix, these two coordinate systems can be transformed into each other
and, employing of $ S_{z_{\theta}} = \cos\theta S_{z}-\sin\theta (\cos\phi S_{x}-\sin\phi S_{y})$,
we can express the eigenstate $|m_s=0\rangle_{\theta}$ in terms of the eigenstates $|0\rangle,|\pm 1\rangle$
of eq. (\ref{hamNV}), i.e. in the lab frame, as
\begin{equation}
\label{optical}
|0\rangle_{\theta} = \cos\theta|0\rangle + \frac{\sin\theta}{\sqrt{2}}
(e^{i\phi}|+1\rangle-e^{-i\phi}|-1\rangle).
\end{equation}
If $\theta$ is large, the eigenstate $|0\rangle$ of the NV center in the laboratory frame as
given by Hamiltonian eq. (\ref{hamNV}) differs significantly from the zero-field eigenstate
$|0\rangle_{\theta}$ of the NV center. Hence optical initialization of randomly oriented NV
centers lead to very different states depending on the orientation of the NV center.

Notice though that for moderate misalignment between the NV center and external magnetic field,
the initialisation of the NV center is well approximated by $|0\rangle$. Indeed, for $\theta<10^\circ$
degrees, $\left|\langle0|0\rangle_{\theta}\right|^2>0.97$ and for $\theta<20^\circ$ we find
$\left|\langle0|0\rangle_{\theta}\right|^2>0.88$ which implies significant polarization along
the quantization axis defined by the external magnetic field. Therefore, for all the orientations
of the NV centers that fall into two spherical sectors whose cone angle is $2\theta$ , i.e.,
the two blue spherical sectors in Fig. \ref{Schematic}a, significant optical polarization can be
achieved. From now on, we define a deviation by $\theta$ to imply that the symmetry axis of the
NV spins falls within these two spherical cones.

\subsection{The main ideas}
\begin{figure}[hbt]
\center
\includegraphics[width=2.8in, height=1.7in]{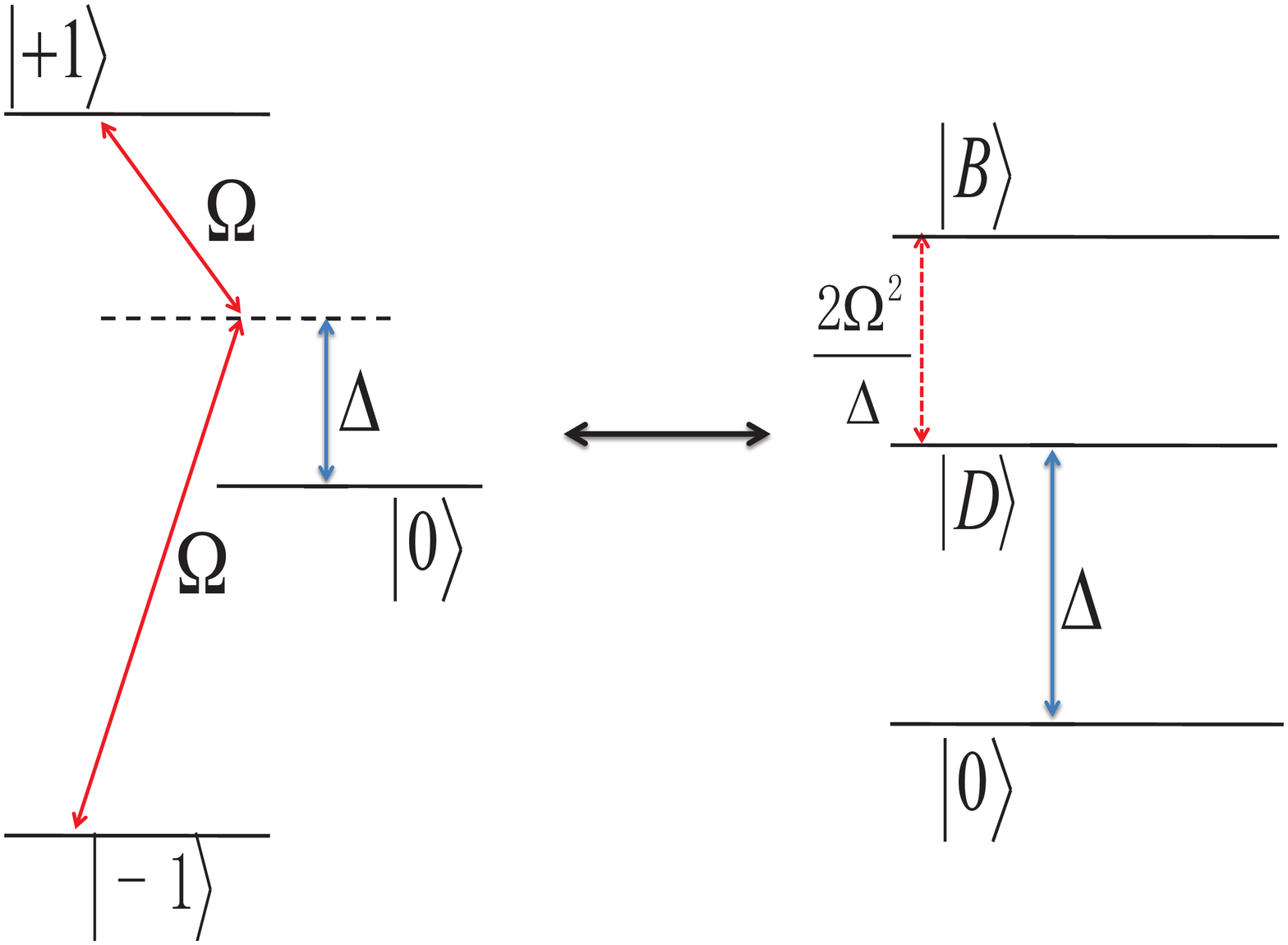}
\caption{The level structure of the three level Hartmann-Hann. A gap is created between the $\vert D \rangle
= \frac{1}{\sqrt{2}}\left( \vert +1 \rangle +  \vert -1 \rangle  \right)$ and the  $\vert B \rangle =
\frac{1}{\sqrt{2}}\left( \vert +1 \rangle -  \vert -1 \rangle  \right)$ states.  This gap is robust to
changes in the one photon detuning ($\Delta$) and is also robust to small changes in the two photon
detuning($\delta$).  Due to the fact that the coupling to the nuclei is via the dressed states($\vert D
\rangle,\vert B \rangle$) and not the bare states the coupling is not decreased due to the off resonant drivings.}
\label{scheme1}
\end{figure}

The main idea that we will develop in this section consists of three main ingredients, namely (i) the
generation of an energy gap between electronic states that is relatively robust with respect to variations
in the relative orientation between the NV center and an external magnetic field and at the same time
(ii) a strong coupling between these electronic states and nuclear spins. Finally, (iii) the integrated
solid effect is employed to achieve additional robustness against imperfections. All this is achieved
by the use of dressed states in a ladder type configuration in which the single quantum transitions are
far detuned and the double quantum transition is nearly resonant (see Fig. \ref{scheme1}).

In such a setting, that is for a detuning on the single quantum transition that is $\Delta(\theta)$,
the effective Rabi frequency and thus the energy splitting on the double quantum transition is proportional
to $\Omega^2/\Delta(\theta)$. In leading order in $\theta$, this suffers a variation that is of the
order of $\Omega^2/\Delta^2\frac{\partial\Delta}{\partial\theta}\theta$. This contrasts with the case of detuned
driving one single quantum transition only. In that case the effective energy gap is proportional to
$\sqrt{\Delta^2 + \Omega^2}$ which, if $\Delta\gg\Omega$, suffers a variation that is of order
$\frac{\partial\Delta}{\partial\theta}\theta$. Therefore the variation in the former is suppressed by
a factor of $\Omega^2/\Delta^2$ compared to the latter. This makes it much easier to achieve and maintain
a Hartmann-Hahn resonance.

The second crucial ingredient is related to the fact that the coupling between the electron spin and
the nuclear spins is mediated by terms of the form $S_z\otimes I_z$. In a far-detuned single quantum
transition the eigenstates involved in the Hartmann-Hahn resonance are $|m_s=0\rangle + \frac{\Omega}{\Delta}
|m_s=+1\rangle$ and $|m_s=+1 \rangle - \frac{\Omega}{\Delta} |m_s=0 \rangle$. As these states are close
to being eigenstates of $S_z$ they hardly couple and in consequence the electron-nuclear coupling is
suppressed by a factor proportional to $\frac{\Omega}{\Delta}$ when compared to the coupling between
states equally weighted superpositions $|m_s=0\rangle \pm |m_s=+1\rangle$ which one would have obtained
for resonant driving. Crucially, thanks to the small detuning on the double quantum transition we obtain
$|m_s=-1\rangle \pm |m_s=+1\rangle$ which lead to a strong coupling between electron and nuclear spins
if a Hartmann-Hahn resonance is realised (see r.h.s of Fig. 3). The latter is relatively robust thanks
to the strong detuning on the single quantum transitions.

So far we have ignored additional energy shifts on the double quantum transition $\delta(\theta)$
which will lead to a loss of the Hartmann-Hahn resonance. In order to confer additional robustness to
our scheme, rather than aiming to maintain a fixed Hartmann-Hahn resonance, we consider an adiabatic
sweep of the external magnetic field on the double quantum transition where start and endpoint of the
sweep are chosen such that intermittent Hartmann-Hahn resonances are guaranteed. This technique, the
integrated solid effect, allows polarization transfer from a large fraction of the NV electron spins
to their surrounding nuclear spins.

Therefore, the resonant double quantum transition that is mediated by far-detuned single quantum
transitions combines robustness with strong electron-nuclear coupling. It is this idea and its
combination with the integrated solid effect that we are going to develop and analyse in detail
for realistic experimental parameters and various imperfections in this section.


\subsubsection{Double quantum transition}

For illustration of the basic idea, let us first consider a system that is composed of an
NV center and a single $^{13}C$ nuclear spin in a nanodiamond of random orientation. In the
laboratory frame the applied magnetic field defines the z-axis, the NV is placed at the origin
of the coordinate system and the $^{13}C$ nuclear spin, placed near the NV center, is situated
at position $\vec{r}$. A microwave (MW) field of frequency $\omega_M$ is applied as off-resonant
drive of the spin transitions $|-1\rangle \leftrightarrow |0\rangle$ and $|0\rangle \leftrightarrow |+1\rangle$
while approximately satisfying a double resonance condition, that is $2\omega_M \cong E_{|+\rangle}
- E_{|-\rangle}$. We chose circular polarization but linear polarization would suffice. The Hamiltonian
of the whole system is then
\begin{eqnarray}
    H &=& \Omega_M(S_x\cos\omega_M t+S_y\sin\omega_M t)+(\gamma_{e}B+\delta(\theta))S_{z}\nonumber \\
    && + D(\theta)S_{z}^{2} + g\Big[\vec{S}\vec{I}-3(\vec{S}\cdot\vec{e}_r)(\vec{I}\cdot\vec{e}_r)\Big]
    +\gamma_{n}BI_{z},
\end{eqnarray}
in which $g=\frac{\mu_0}{4\pi}\frac{\gamma_e\gamma_n}{r^3}$ with $r=|\vec{r}|$ denoting the
distance from the NV spin to the nuclear spin and $\vec{e}_r = \vec{r}/r$. $\vec{I}$ is
spin-$\frac{1}{2}$ vector operator of the nuclear spin with the gyromagnetic ratio $\gamma_n$
and $\Omega_M=\sqrt{2}\Omega$ is the Rabi frequency of the driving field. Assuming a point-dipole
interaction and neglecting the contact term we obtain
\begin{eqnarray}
\label{point dipole}
    H_e &\simeq& \Omega_M(S_x\cos\omega_M t+S_y\sin\omega_M t) \\ \nonumber
    &&+(\gamma_{e}B+\delta(\theta))S_{z}+D(\theta)S_{z}^{2} + \gamma_{n}BI_{z}\\ \nonumber
    && - gS_z\Big[3e^z_r(e^x_rI_x+e^y_rI_y)+(3(e^z_r)^2-1)I_z\Big]. \nonumber
\end{eqnarray}
In the interaction picture with respect to $H_{0}=\omega_M S_{z}$, the Hamiltonian is given by
\begin{eqnarray}
    H' &=& \Omega(|-1\rangle\langle0|+|0\rangle\langle +1|+h.c.)\nonumber \\
    && +\gamma_{n}BI_{z} +D(\theta)S_{z}^{2}+\Delta S_z+S_{z}\cdot\mathbf{A}\cdot\vec{I},
\end{eqnarray}
in which $\mathbf{A}=g\sqrt{1+3(e^z_r)^2}\vec{h}=A\vec{h}$ is the hyperfine interaction tensor
for the nuclear spin, with $\vec{h}$ determined by $e_r$: $h_x=3e^x_re^z_r/\sqrt{1+3(e^z_r)^2}$,
$h_y=3e^y_re^z_r/\sqrt{1+3(e^z_r)^2}$, and $h_z=(3(e^z_r)^2-1)/\sqrt{1+3(e^z_r)^2}$. The detuning $\Delta$
is given by
\begin{equation}
    \Delta= \gamma_{e}B+\delta(\theta)-\omega_M,
\end{equation}
which is a function of the NV orientation due to the second order energy correction $\delta(\theta)$.

Since the nuclear coordinate system is arbitrary, we can redefine it by following the method in Ref.~\cite{London}.
The z-axis is chosen to coincide with the external magnetic field, and we define the y-axis to be perpendicular
to both magnetic field direction and $\vec{h}$. In the new nuclear coordinate system, the Hamiltonian is given by
\begin{eqnarray}
    \label{FullHam}
    H'' &=& \Omega(|-1\rangle\langle0|+|0\rangle\langle +1| + h.c.)+ D(\theta)S_{z}^{2} +\Delta S_{z} \nonumber \\
    && +\gamma_{n}BI_{z'}+S_{z}\cdot(a_{x'}I_{x'} + a_{z'}I_{z'})
\end{eqnarray}
where $I_z = I_{z'}$ and $a_{z'}$ and $a_{x'}$ are the elements of the secular and
pseudosecular hyperfine interactions, respectively.

In what follows we assume that (i) $|D(\theta)|\gg\Delta$ which implies that the microwave driving
field tends to be far detuned from the single quantum transitions $|-1\rangle \leftrightarrow |0\rangle$
and $|0\rangle \leftrightarrow |+1\rangle$ and (ii) $|D(\theta)|\gg\Omega$ such that the Rabi frequency
is significantly smaller than the detuning (see Fig. \ref{Schematic}b). This will allow us to remove
level $|0\rangle$ adiabatically from the dynamics to achieve a simplified effective Hamiltonian eq. (\ref{H plus}).


To this end, we consider the dominant parts of eq. (\ref{FullHam}) by neglecting the nuclear spins as well
as $\Delta S_z$. In the basis $\{|+1\rangle,|0\rangle,|-1\rangle\}$ we therefore consider
\begin{equation}
    H_{NV}	=	\left(\begin{array}{ccc}
        D(\theta) & \Omega & 0\\
        \Omega & 0 & \Omega\\
        0 & \Omega & D(\theta)
    \end{array}\right).
\end{equation}
The eigenstates and eigenenergies of $H_{NV}$ are
\begin{eqnarray}
    |\mu_{\mp}\rangle &=& \frac{1}{\sqrt{2+X_{\pm}^{2}}}(|+1\rangle+|-1\rangle- X_{\pm}|0\rangle), \\
    \omega_{\mu_{\pm}} &=& \frac{1}{2}[D(\theta)\pm\sqrt{8\Omega^{2}+D^{2}(\theta)}]
\end{eqnarray}
and
\begin{eqnarray}
    |\lambda\rangle &=& \frac{1}{\sqrt{2}}(|+1\rangle-|-1\rangle),  \\
    \omega_{\lambda} &=& D(\theta)
\end{eqnarray}
with
\begin{eqnarray*}
    X_{\pm}(D(\theta),\Omega) &=& \frac{D(\theta)\pm \sqrt{8\Omega^{2}+D^2(\theta)}}{2\Omega}.
\end{eqnarray*}
For $|D(\theta)|\gg\Omega$ we have two cases: $D(\theta)>0$ leads to $X_{-}(D(\theta), \Omega)\sim0$
and $X_{+}(D(\theta), \Omega) \gg 1$, while $D(\theta)<0$ results in $X_{+}(D(\theta), \Omega)\sim0$
and $X_{-}(D(\theta), \Omega) \gg 1$.

In both cases, two of the eigenstates are approximately given by $|\pm\rangle=\frac{1}{\sqrt{2}}(|-1\rangle\pm|+1\rangle)$
and form an effective two-level system while the state $|0\rangle$ does not participate in the dynamics
because for $|D(\theta)|\gg\Omega$ it is far detuned. We will now focus attention on these two states
$|\pm\rangle$ and write the Hamiltonian eq. (\ref{FullHam}) in the subspace spanned by $\{|+\rangle,|-\rangle\}$
to find
%
%
%
\begin{eqnarray}
\label{H plus}
    H_{\pm}&\simeq&\pm\Omega_{eff}\sigma_z+2\Delta\sigma_{x}+\gamma_{n}BI_{z'}+2\sigma_{x}(a_{x'}I_{x'}+a_{z'}I_{z'})\nonumber\\
\end{eqnarray}
in which $H_+$ and $H_-$ corresponding to the cases $D(\theta)>0$ and $D(\theta)<0$, respectively.
Here
\begin{displaymath}
    \Omega_{eff}=\frac{1}{2}[-|D(\theta)|+\sqrt{8\Omega^{2}+D^{2}(\theta)}]
\end{displaymath}
and $\sigma_z=\frac{1}{2}(|+\rangle\langle+|-|-\rangle\langle-|)$, $\sigma_x=\frac{1}{2}(|+\rangle\langle-|+|-\rangle\langle+|)$,
$I_{z'}=\frac{1}{2}(|\uparrow\rangle\langle \uparrow|-|\downarrow\rangle\langle\downarrow|)$
and $I_{x'}=\frac{1}{2}(|\uparrow \rangle\langle\downarrow|+|\downarrow\rangle\langle\uparrow|)$.
Furthermore $|\downarrow\rangle$ and $|\downarrow\rangle$ denote the ground and excited states of the nuclear spin.

For the final step, we now restrict attention to the case where $\theta\in[0^\circ,20^\circ]$ in which case
$D(\theta)>0$ and we consider Hamiltonian $H_{+}$.
In matrix notation, the electronic part of the Hamiltonian eq. (\ref{H plus}) then takes the form
\begin{equation}
    \label{H detuning}
    H_{\Delta}	=	\left(\begin{array}{cc}
    \frac{\Omega_{eff}}{2} & \Delta\\
    \Delta & -\frac{\Omega_{eff}}{2}
    \end{array}\right).
\end{equation}
The eigenstates and eigenenergies are
\begin{eqnarray}
    |\chi_-\rangle &=& \cos\frac{\zeta}{2}|-\rangle-\sin\frac{\zeta}{2}|+\rangle \nonumber\\
    |\chi_+\rangle &=& \cos\frac{\zeta}{2}|+\rangle+\sin\frac{\zeta}{2}|-\rangle,\nonumber \\
    \omega_{eff} &=& \pm\sqrt{\Delta^2+\frac{\Omega_{eff}^2}{4}},
\end{eqnarray}
in which $\arctan\zeta=\frac{-\Delta}{\Omega_{eff}/2}$, $\Delta$ and $\Omega_{eff}$ are dependent
on the angle $\theta$ between NV orientations and applied magnetic field. Defining Pauli operators
$\sigma_{\tilde x},\sigma_{\tilde y},\sigma_{\tilde z}$ in the eigenbasis of $H_{\Delta}$ we find
\begin{eqnarray}
    H_+ &=& 2\omega_{eff} \sigma_{\tilde z}+\gamma_{n}BI_{z'} \\ \nonumber
    && +2(\sigma_{\tilde x}\sin\varphi + \sigma_{\tilde z}\cos\varphi)\cdot(a_{x'}I_{x'}+a_{z'}I_{z'}),
 \end{eqnarray}
in which the angle $\varphi$ is determined by
\begin{displaymath}
    \sin\varphi=\frac{\Omega_{eff}}{\sqrt{4\Delta^2+\Omega_{eff}^2}}. \nonumber
\end{displaymath}

The Hamiltonian $H_+$ can be simplified further under two main assumptions, namely (i) $a_{x'}\ll\gamma_{n}B$,
that is for weakly coupled nuclear spins and (ii) $|\frac{\gamma_{n}B}{2}-\omega_{eff}| \ll a_{x'}$
that is we satisfy a Hartmann-Hahn condition. Then the polarization transfer dynamics is described by
\begin{eqnarray}
    \label{Htrans}
    H_{trans} &=& 2\omega_{eff}\sigma_{\tilde z}+\gamma_{n}BI_{z'} + 2a_{z'}\cos\varphi \sigma_{\tilde z} I_{z'}\nonumber\\
    && + \frac{a_{x'}\sin\varphi}{2}(|\chi_+, \downarrow\rangle\langle \chi_-, \uparrow| + h.c.)
\end{eqnarray}
where $2a_{z'}\cos\varphi \sigma_{\tilde z} I_{z'}$ does not affect the flip-flop interaction.

Let us briefly highlight the key difference of this off-resonant driving in comparison to previously introduced
polarization schemes, namely its relatively high robustness with respect to the angle $\theta$ between the external
magnetic field and the axis of the NV center. If we were to follow the scheme of Ref.~\cite{London}, even for
$\theta=1^\circ$, the detuning of the MW frequency from the NV electronic resonance exceeds $(2\pi)1$ MHz, which
prevents efficient polarization transfer (for $\theta =20^\circ$ we find $(2\pi)500$ MHz detuning). In our scheme,
for as long as $D(\theta) \gg \Omega$, the energy difference between the states $|-\rangle$ and $|+\rangle$
scales as $2\Omega^2/D(\theta)$ and so does the effective Rabi frequency $\Omega_{eff}$. For deviation
$\theta\in[0^\circ,20^\circ]$, if $\Omega=(2\pi)65$ MHz, the effective Rabi frequency is roughly within
$[(2\pi)3MHz, (2\pi)3.6MHz]$, which narrows the range of the detuning of the resonant frequency.

\subsubsection{Integrated solid effect}
\label{Integrated solid effect}

In the discussion so far, we have assumed that the electron spin of the NV center is held continuously
at a Hartmann-Hahn resonance with a specific target nuclear spin to achieve polarization transfer. In
practice, however, this will only capture a small fraction of the NV centers and nuclear spins as the
resonance condition will depend on $\Delta = \gamma_e B + \delta(\theta) - \omega_M$ which in turn is
a function of the angle $\theta$ via $\delta(\theta)$ (Fig. \ref{distribution}b). In the range
$\theta\in[0^\circ,20^\circ]$, we have $\delta(\theta)\in[0,(2\pi)45MHz]$, as shown in Fig. \ref{distribution}b
which can lead to a violation of the Hartmann-Hahn resonance for a large fraction of the NV centers.
In order to achieve polarization transfer for a larger fraction of the NV centers, we will use our
ability to control the strength of the externally applied magnetic field or the frequency of the applied
microwave field to implement a sweep of $\Delta$.

In order for us to understand qualitatively the conditions that such a sweep has to satisfy, we
briefly examine the energy level diagram of a system consisting of one electron and one nuclear spin
as described by eq. (\ref{Htrans}). This reveals that there are two resonance points $A_1$ and $A_2$
for $\Delta=\pm\Delta_{HH}$ owing to the symmetry of the Hamiltonian. In a fully adiabatic sweep,
polarization would exchange twice and result in a vanishing net polarization transfer. Hence the sweep
rate needs to be chosen such that around the $A_i$ the system becomes non-adiabatic for the two branches
including $|\chi_+,\downarrow\rangle$ and $|\chi_-,\uparrow\rangle$ as this will lead to an approximately
$50\%$ probability for polarization transfer. At the same time the sweep has to remain adiabatic with respect
to the energy gap for transitions between the $\{|\chi_+,\downarrow\rangle, |\chi_-,\uparrow\rangle\}$
and the $\{|\chi_+,\uparrow\rangle , |\chi_-,\downarrow\rangle\}$ manifold as transitions between these
two manifolds, induced by non-adiabaticity or in fact dephasing events, would lead to depolarization.
These requirements set some limitations on the sweep rate that are, fortunately, not too stringent.

The quasi-adiabatic sweep described above realises an instance of the so-called integrated solid effect
(ISE)~\cite{Henstra1} and can achieve polarization transfer between a large fraction of the NV spins and
nuclear spins. Therefore, magnetic control of the nuclear spin, and efficient polarization exchange, become
possible even for relatively large angles between NV-center axis and external magnetic field. In the
following we will examine this idea in more detail.


Let us now examine two regimes of the quasi-adiabatic transfer described above, namely (i) those
parts of the sweep parameters that are far from the Hartmann-Hahn resonance points $A_1$ and $A_2$ which
will lead to an upper bound on the sweep rate and (ii) the behaviour around the Hartmann-Hahn
resonances which will provide lower bounds on the sweep rate.

\begin{figure}
\center
\includegraphics[width=2.8in]{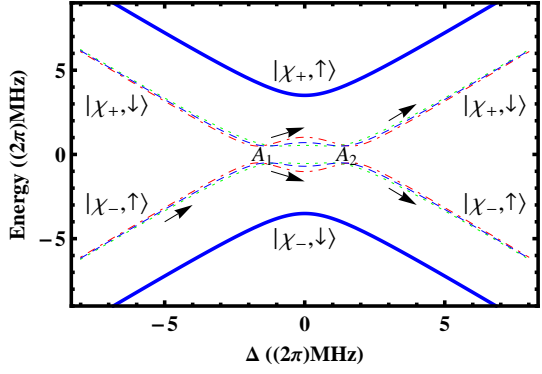}
\caption{Eigenenergies of eq. (\ref{Htrans}) for $B=0.36$ T in which $|\frac{a_{x'}\sin\varphi_l}{2}|=(2\pi)0.5$ MHz,
$\Omega_{eff}=(2\pi)2.2$ MHz (the dot-dashed red line), $\Omega_{eff}=(2\pi)3$ MHz (the dotted green line),
$\Omega_{eff}=(2\pi)3.6$ MHz (the dashed blue line). The level diagram shows the progression of the eigenvalues
of the Hamiltonian during the ISE sweep. Energies of the states $|\chi_+,\uparrow\rangle$ and $|\chi_-,\downarrow\rangle$
(solid blue line) for $\Omega_{eff}=(2\pi)2.2$ MHz are included. Transitions from states represented by solid blue
lines to state represented by dashed lines may lead to depolarization of the nuclear bath.}
\label{Eigenenergies}
\end{figure}

For regime (i), as the coupling between electron and nuclear spin is negligible, it suffices to
examine the electronic part of eq. (\ref{H plus}), that is
\begin{equation}
    H'_{\Delta}	=	\left(\begin{array}{cc}
    \frac{\Omega_{eff}}{2} & \Delta(t)\\
    \Delta(t) & -\frac{\Omega_{eff}}{2}
    \end{array}\right).
\end{equation}
in which $\Delta(t)= \gamma_{e}B-\delta(\theta)-\omega_M$. For simplicity we assume the detuning
to vary at a constant rate $v>0$, i.e. $\Delta(t)=\Delta(t_i)+vt$, where $\Delta(t_i)$ is the
initial detuning at the start of sweep. According to the quantum adiabatic condition we require,
at all times, that
\begin{equation}
    \frac{\langle E_{1m}(t)|\dot{E}_{1n}(t)\rangle}{|E_{1m}(t)-E_{1n}(t)|}\gg1, \ \ \ \ \ m\neq n,
\end{equation}
where $E_{1m}(t)$ and $E_{1n}(t)$ are the instantaneous eigenvalues, $|E_{1m}(t)\rangle$ and
$|E_{1m}(t)\rangle$ are the eigenstates of effective Hamiltonian $H_{\Delta}$, respectively.
Then the condition for the sweep to be adiabatic is given by
\begin{equation}
    \label{ad condition}
    \frac{\Omega_{eff}^2}{|v|}\gg1.
\end{equation}
Adiabatic evolution implies that the initial eigenstate of this Hamiltonian $H_{\Delta}$ will remain
close to the instantaneous eigenstate at any time during the sweep. In particular, during approach
to the Hartmann-Hahn resonances and indeed during the entire sweep there will be no transitions
of the type $|\chi_-,\downarrow\rangle$ to $|\chi_+,\downarrow\rangle$ which, as is easily seen by
examination of Fig. \ref{Eigenenergies}, would have a depolarizing effect in the subsequent sweep.

To gain a feeling for the upper limit on the sweep rates that this implies for typical experimental
parameters suppose as in the previous section that $\Omega=(2\pi)65$ MHz so that in the range
$\theta\in[0^\circ,20^\circ]$ we find $\Omega_{eff}$ in the range $[(2\pi)3 MHz,(2\pi)3.6 MHz]$.
This in turn implies that $v<(2\pi)10$ MHz$/\mu$s guarantees that the adiabatic condition is satisfied.
As shown in Fig. \ref{distribution}b in this range we have $\delta(\theta)\in[0,(2\pi)45MHz]$. Suppose
a sweep rate $v=(2\pi)6$ MHz$/\mu$s, then no more than 10 $\mu$s are required to cover all NV centers
within $\theta\in[0^\circ,20^\circ]$. Experimentally such sweep rates are easily obtained using arbitrary
wave form generators.

Let us now proceed to discuss case (ii), namely the sweep around the Hartmann-Hahn resonances which
will provide lower bounds in the sweep rate. Near the resonant points, the effective Hamiltonian
eq. (\ref{Htrans}) in the subspace spanned by $\{|\chi_+,\downarrow\rangle, |\chi_-,\uparrow\rangle\}$
is well approximated by the matrix form
\begin{equation}
    H^{l}_{mat}	=	
    \left(\begin{array}{cc}
        (-1)^{l}v(t-t_l)\cos\varphi_l & \frac{a_{x'}\sin\varphi_l}{2}\\
        \frac{a_{x'}\sin\varphi_l}{2} & (-1)^{l+1}v(t-t_l)\cos\varphi_l
    \end{array}\right).
\end{equation}
Here $l=1,2$ corresponds to points $A_1$ and $A_2$, and near these two points we make use of the
expansion
\begin{displaymath}
    \omega_{eff}(t)= \omega_{eff}(t_l) + \frac{\partial\omega_{eff}(t)}{\partial t}|_{t_l} (t-t_l)
    = (-1)^{l}v(t-t_l)\cos\varphi_l.
\end{displaymath}
According to Landau-Zener (LZ) theory~\cite{Zener}, transitions are possible between two approaching
levels as a control parameter is swept across the point of minimum energy splitting. The asymptotic
probability of a LZ-tunneling transition is given by
\begin{equation}
    P_{LZ} = e^{-2\pi\mu}
\end{equation}
where, by virtue of the Hartmann-Hahn condition $\gamma_n B = \sqrt{4\Delta^2 + \Omega_{eff}^2}$ at resonant pints $A_1$ and $A_2$, we find
\begin{eqnarray*}
  \mu&=&\frac{(\frac{a_{x'}\sin\varphi_l}{2})^2}{2|v|\cos\varphi_l}
  =\frac{\Omega_{eff}^2a_{x'}^2}{8|v|(\gamma_nB)\sqrt{(\gamma_nB)^2-\Omega_{eff}^2}}.
\end{eqnarray*}
This enters the unitary transformation
\begin{equation}
    N^{l}	=	
    \left(\begin{array}{cc}
        \sqrt{1-P_{LZ}}e^{i\tilde{\varphi}_s}& -\sqrt{P_{LZ}} \\
          \sqrt{P_{LZ}}& \sqrt{1-P_{LZ}}e^{-i\tilde{\varphi}_s}
    \end{array}\right).
\end{equation}
that is taking place when traversing the avoided crossing. Here $\tilde{\varphi}_s=\varphi_s-\pi/2$ is
due to the Stokes phase $\varphi_s$.

\begin{figure}[b]
\center
\includegraphics[width=2.8in]{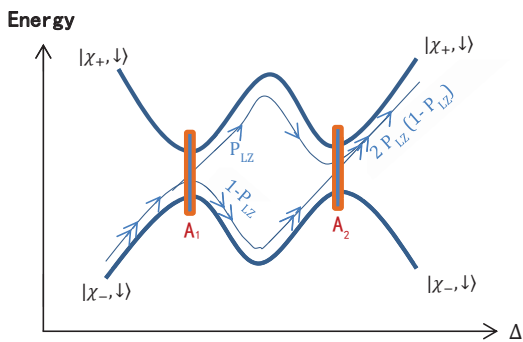}
\caption{Double-passage transition.  A schematics of a pair of the adiabatic energy levels as in
Fig. \ref{Eigenenergies}. At each crossing there is a probability of $P_{LZ}$ for a Landau-Zener
transitions. The lines with one (two) arrows show the two paths where the transition to the upper
level happens during passage of the first (second) resonance point. The average probability for
a polarization transfer at the end of the completed passage is given by $2P_{LZ}(1 -P_{LZ})$.}
\label{double}
\end{figure}

\begin{figure*}[t]
\center
\includegraphics[width=2.3in]{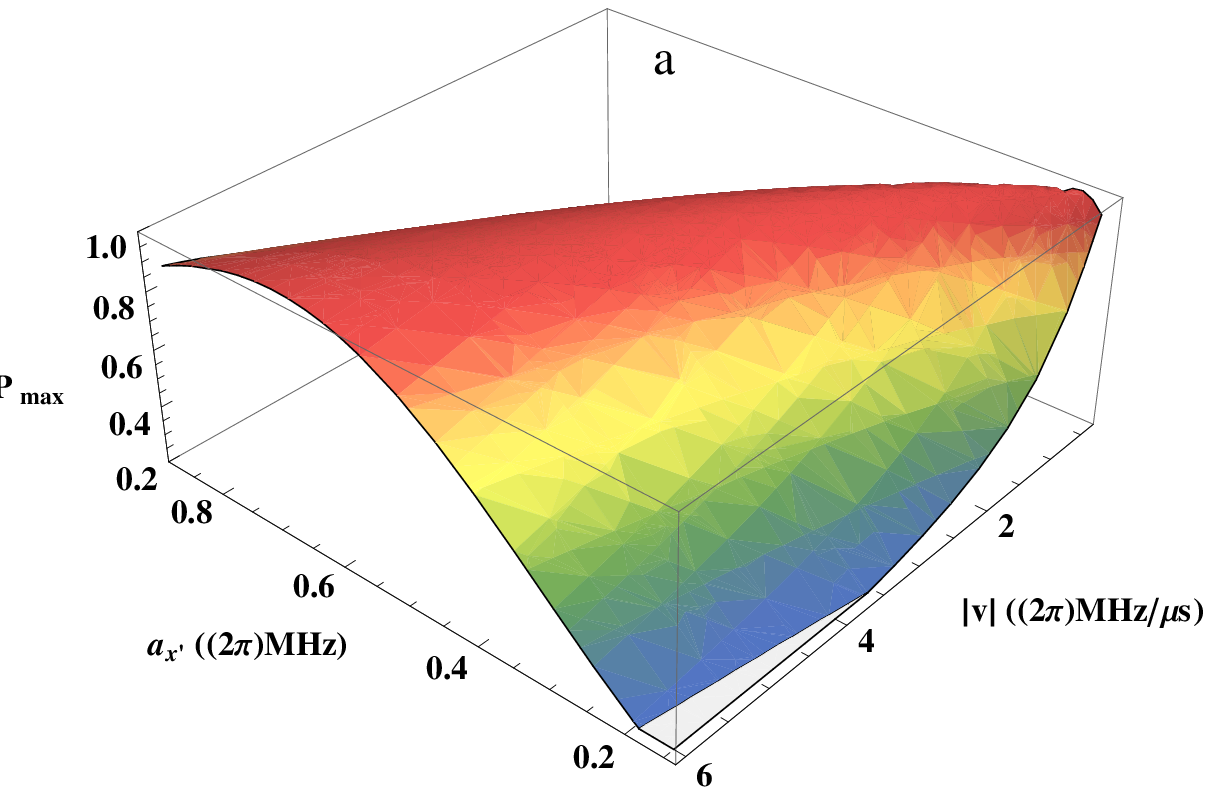}
\includegraphics[width=2.3in]{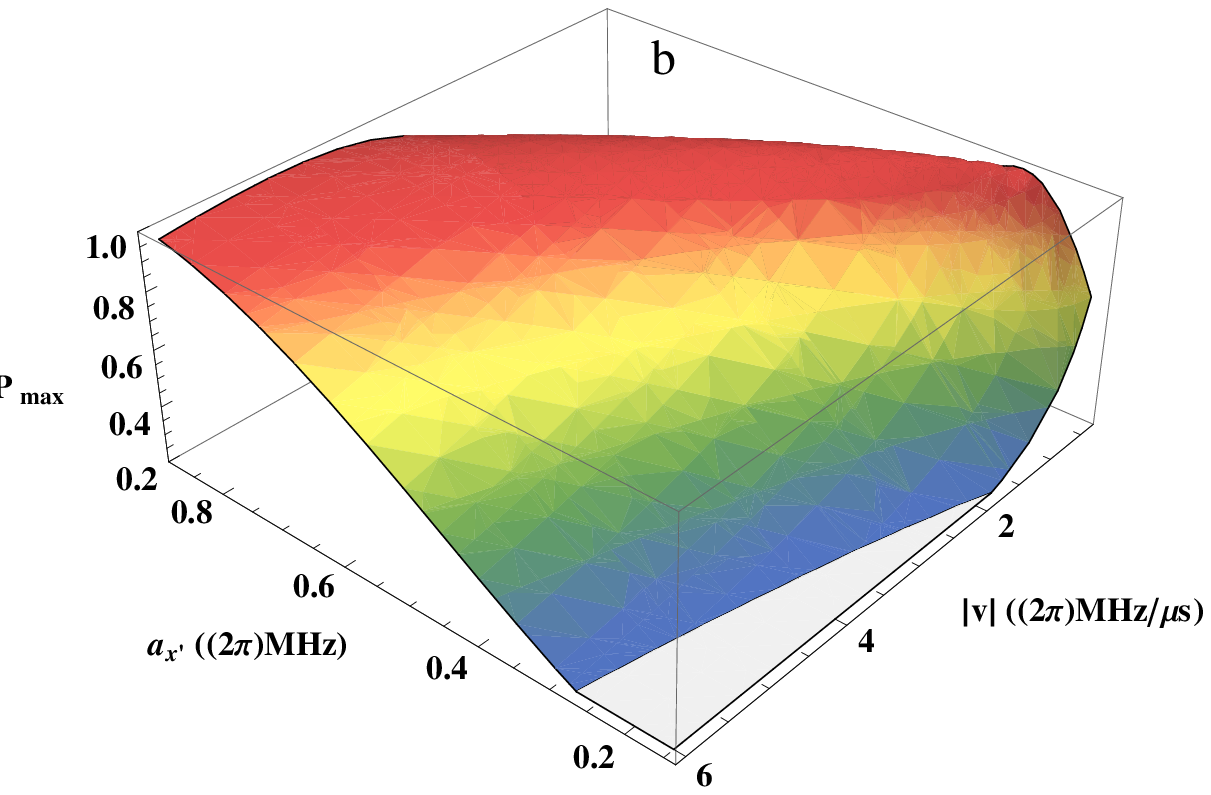}
\includegraphics[width=2.3in]{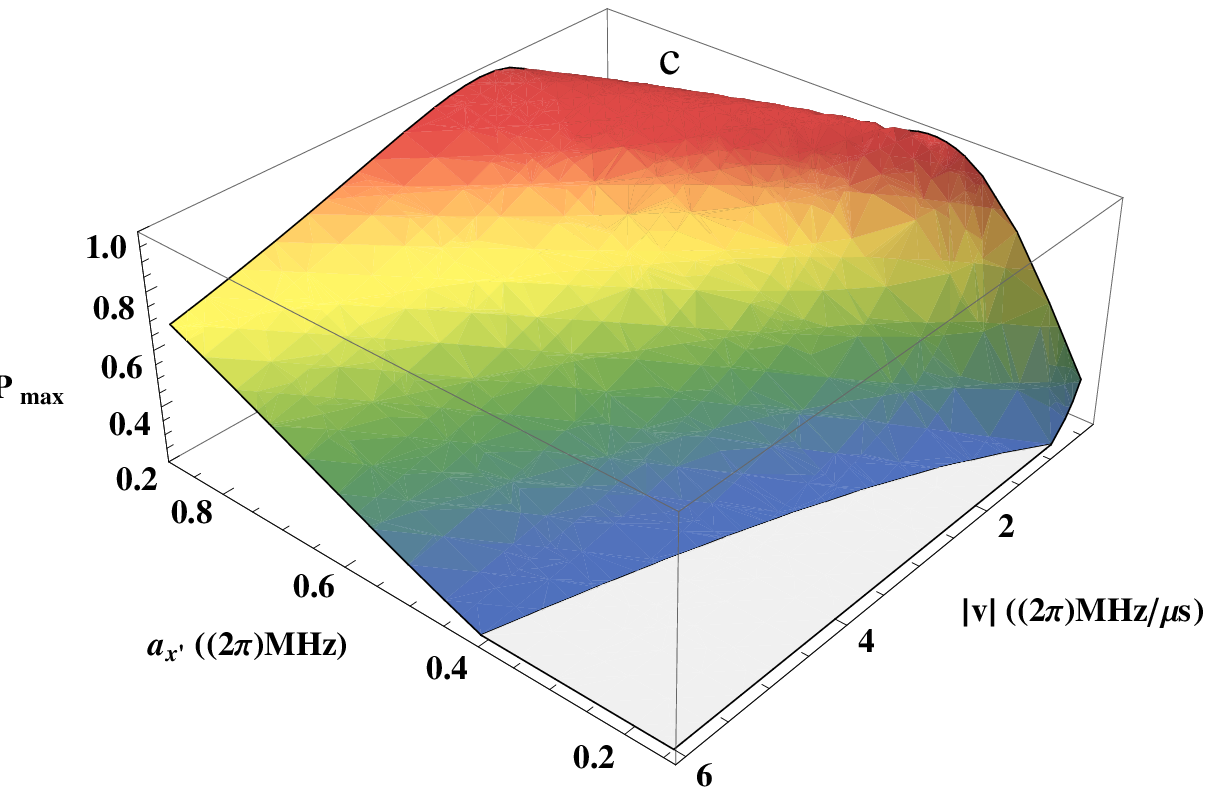}
\caption{The maximal polarization transfer $P_{max}$ as a function of the effective coupling $\Omega_{eff}$.
a, $\Omega_{eff}=(2\pi)3.6$ MHz£» b, $\Omega_{eff}=(2\pi)3$ MHz;  c, $\Omega_{eff}=(2\pi)2.3$ MHz.}
\label{pmax}
\end{figure*}

The full sweep takes the system from $\Delta\ll 0$ to $\Delta\gg 0$ across both $A_1$ and $A_2$. At each
resonance point polarization transfer is possible between the electron and the nuclear spins~\cite{Henstra2}.
Consider a sweep starting at $\Delta(0)\ll -\Omega_{eff}$ for which the electron spin is initialized in state
$|\chi_-\rangle$ and the adiabatic condition eq. (\ref{ad condition}) is satisfied. If the nuclear spin is
initially found in state $|\downarrow\rangle$, the electron-nuclear system will start and finish the sweep in
state $|\chi_-, \downarrow\rangle$, i.e. the nuclear polarization is unchanged. If on the other
hand the nuclear
spin is initially found in state $|\uparrow\rangle$, the evolution of the system will start in $|\chi_-, \uparrow\rangle$.
In this case the probability for experiencing a polarization transfer at the end of the full sweep, that is
finding the system in state $|\chi_+,\downarrow\rangle$, is given by
\begin{eqnarray}
\label{Pmax}
   P&=&P_{max}\sin^2[\Phi_{St}] \nonumber \\
   &=&  4P_{LZ}(1-P_{LZ})\sin^2[\Phi_{St}] .
\end{eqnarray}
Here $\Phi_{St}$ is determined by the initial phase of the system and the phases acquired during the adiabatic
evolution and the non-adiabatic transitions. We do not discuss these dynamical phase shifts in detail as they
depend on details of the evolution and orientations that vary from NV center to NV center and it is therefore
reasonable to assume that $\Phi_{St}$ is random~\cite{Henstra3,Henstra4}. Hence, when averaged over many sweeps,
the average polarization transfer probability is given by
\begin{eqnarray*}
    \overline{P} = 2P_{LZ}(1-P_{LZ}),
\end{eqnarray*}
as shown in Fig. \ref{double}.
The value of $P_{LZ}$  and hence the maximum $\overline{P} = 2P_{LZ}(1-P_{LZ})$ is determined by the hyperfine
coupling strength $a_{x'}$, the rate of the ISE sweep $v$, and to a lesser extent on the strength of the applied
microwave field.

It is straightforward to see that $P_{max}$ and hence $\overline{P}$ vanish for both $v\gg0$ and $v\rightarrow0$,
corresponding to the diabatic and adiabatic sweep. The maximum value is achieved for $P_{LZ}=1/2$ and the parameters
for achieving this maximum are quite flexible, see Fig. \ref{pmax}. $P_{max}$ remains high across a wide range
of sweep rates as shown in Fig. \ref{pmax}a and b. Note furthermore, that for a given sweep rate we obtain
efficient polarization transfer to the nuclear spins in a broad range of different microwave Rabi frequencies,
see Fig. \ref{pmax}. For example, for $\Omega_{eff}=(2\pi)3.6$ MHz, a speed $v=(2\pi)6$ MHz$/\mu$s can obtain
$P_{max}>0.2$ in the coupling range $a_{x'}\in[(2\pi)0.18,(2\pi)0.9]$MHz. Furthermore, a slower sweep gives a better
polarization transfer for a smaller effective Rabi frequency and a weaker coupling strength between the electron
spin and the nuclear spin. In order to polarize the nuclear spins with larger distance from the NV spin, one can
slow down the ISE sweep, or increase the Rabi frequency of the microwave field, as shown in Fig. \ref{pmax}c.

\subsection{Preparation of the initial state}
So far we have assumed the preparation of a specific initial state in the polarization
sequence to be achieved with perfect fidelity. However, due to the random orientations of the NV centers,
optical pumping leads to the initialization of the electron spin of the NV center in a
wide range of different states. Furthermore, due to the broad distribution of zero-field
splittings of the NV center in randomly oriented nanodiamonds, any applied microwave field
will experience an uncertain and potentially large detuning from the electronic resonance.
This presents challenges to the preparation of the electron spin of the NV center in the
initial state that is required for our polarization scheme. We address these challenges
with an adiabatic sweep of the frequency of the microwave field which we will show to
deliver robust and rapid state preparation.

Let us consider as an example the initialization of the NV center in the state $|-1\rangle$.
For moderate $\theta$ optical pumping will lead to a state that possesses a significant overlap
with $|0\rangle$. In order to map this state to the target $|-1\rangle$, we make use of the
fact that in the relatively high magnetic fields that we are considering (e.g. B$\cong 0.36$T)
the energy gap on the $|-1\rangle\leftrightarrow|0\rangle$ transition is very different from
the energy gap of the $|0\rangle\leftrightarrow|+1\rangle$ transition. Hence it is possible
to use a circular polarized microwave driving field
\begin{eqnarray}
    H_{dr} = \Omega_{M-}(S_x\cos\omega_- t + S_y\sin\omega_- t),
\end{eqnarray}
in which $\Omega_{M-}=\sqrt{2}\Omega_-$ is the Rabi frequencies of the MW field, with frequency $\omega_-$
to drive dominantly the $|-1\rangle\leftrightarrow|0\rangle$ transition. The resulting effective
two-level system consisting of the states $|0\rangle$ and $|-1\rangle$ can be described in the matrix
representation by
\begin{equation}
    H_{0,-1}	=	
    \left(\begin{array}{cc}
        \Delta_{MW}(t)/2 & \Omega_-\\
        \Omega_- & -\Delta_{MW}(t)/2
    \end{array}\right),
\end{equation}
in which $\Delta_{MW}(t)=\gamma_eB+\delta(\theta)+ D(\theta)-\omega_-(t)$. Now we assume that $\omega_-(t)$
experiences a constant rate sweep such that $\Delta_{MW}(t)=\Delta_{MW}(t_i) + \dot{\Delta}_- (t-t_i)$.
We set $t_i=0$ for simplicity. The condition for the sweep to be adiabatic is then
\begin{equation}
    \frac{8\Omega_-^2}{|\dot{\Delta}_-|}\gg 1.
\end{equation}
Adiabatic evolution implies that the initial eigenstate of this Hamiltonian $H_{0,-1}$ will remain close
to the instantaneous eigenstate at any time during the sweep. The eigenstates and eigenenergies are
\begin{eqnarray}
    |E^\pm(t)\rangle &=& \zeta^{\mp}_1(t)|-1\rangle\pm\zeta^{\pm}_1(t)|0\rangle,\nonumber \\
    E^\pm(t) &=& \pm\sqrt{(\frac{\Delta_{MW}(t)}{2})^2 + \Omega_{-}^2},
\end{eqnarray}
with $\zeta^{\pm}_1=\sqrt{\frac{E^+(t)\mp\Delta_{WM}(t)}{2E^+(t)}}$. If at the start of the sweep
$\Delta_{WM}(0) < 0$ and $|\Delta_{WM}(0)|\gg 2\Omega_-$ and the electron spin is prepared in the
state $|0\rangle$, the state will evolve along the path $|E^+(t)\rangle$ to end in the state $|-1\rangle$.

As an example, consider a Rabi frequency $\Omega_-=(2\pi)20$ MHz and $\theta\in[0^\circ,20^\circ]$.
Then a sweep from $\Delta_{WM}(t_i)>(2\pi)160$ MHz to $\Delta_{WM}(t_f) < -(2\pi)160$ MHz includes
all NV centers in the sweep. For $|\omega_1(t_f)-\omega_1(t_i)| = \dot{\omega}_1t_f\simeq (2\pi)870$ MHz
we require $t_f\gg[\omega_1(t_f)-\omega_1(t_i)]/8\Omega_-^2 = 0.04$ $\mu$s to achieve adiabaticity. Hence
very high fidelity preparation of the state $|-1\rangle$ can be achieved within $0.4$ $\mu$s which is
faster than the optical polarization cycle.

\begin{figure}[b]
\center
\includegraphics[width=3.2in]{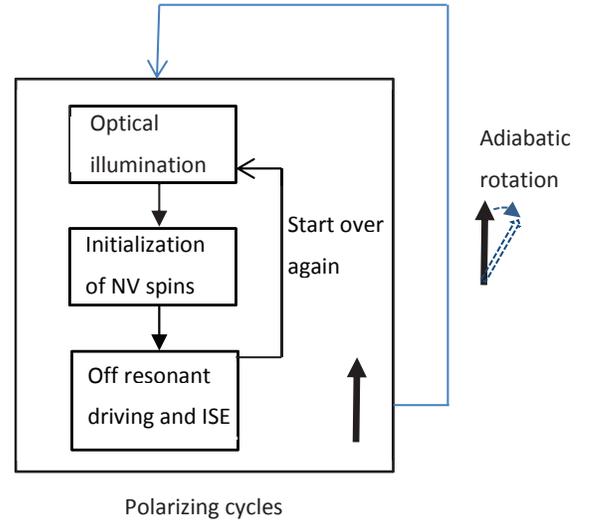}
\caption{A simple flow diagram of the entire polarization scheme: For a given direction of magnetic field,
the large box includes numerous polarization cycles (each polarization cycle includes optical initialization
of the NV spin, adiabatic preparation of the state $|-1\rangle$ and then polarization transfer by matching
a H-H resonance condition by using off-resonant microwave driving and the ISE technique). After the adiabatic
rotation of the magnetic field (The thick black arrow represent the magnetic field orientation which is rotated
to another direction denoted by the dashed blue arrow) we repeat our polarizing cycles to achieve the polarization
of additional nuclear spins.}
\label{flow1}
\end{figure}

\begin{figure*}[t]
\center
\includegraphics[width=2.3in]{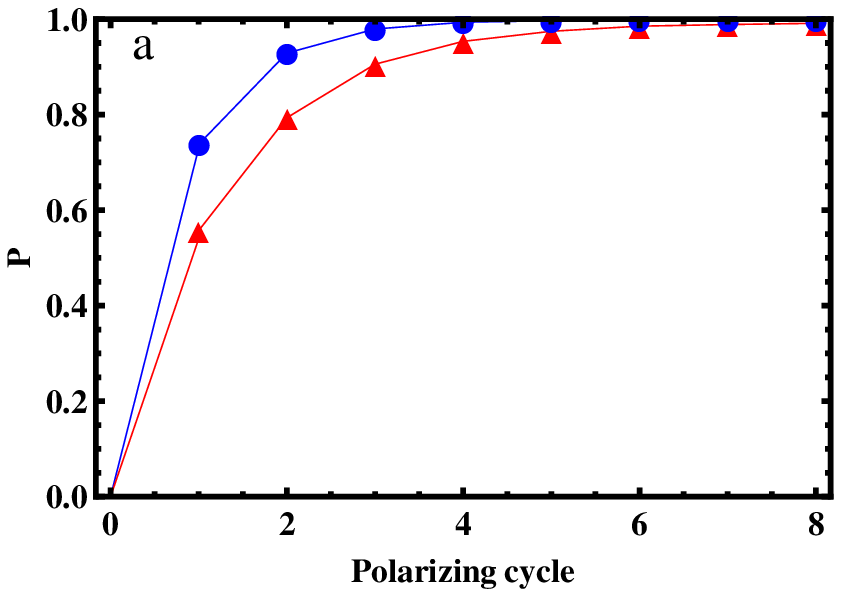}
\includegraphics[width=2.3in]{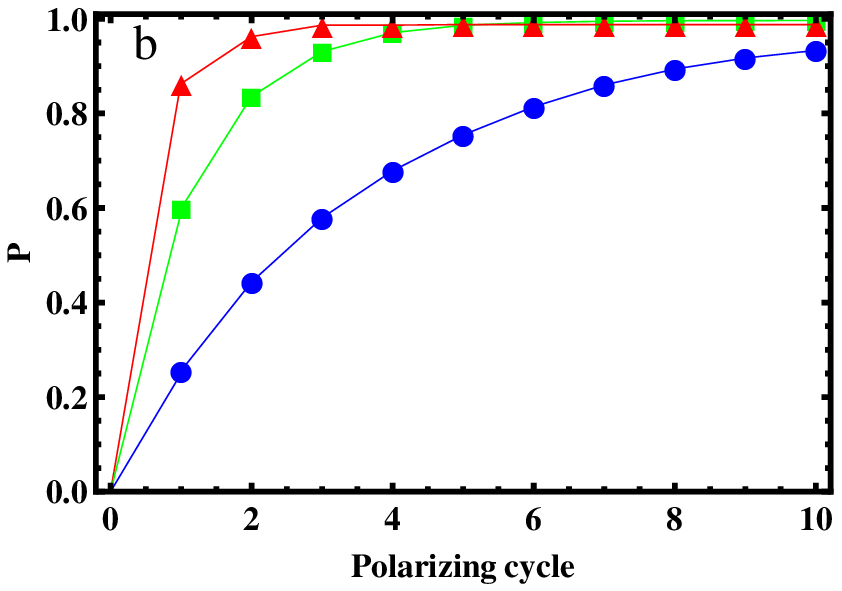}
\includegraphics[width=2.3in]{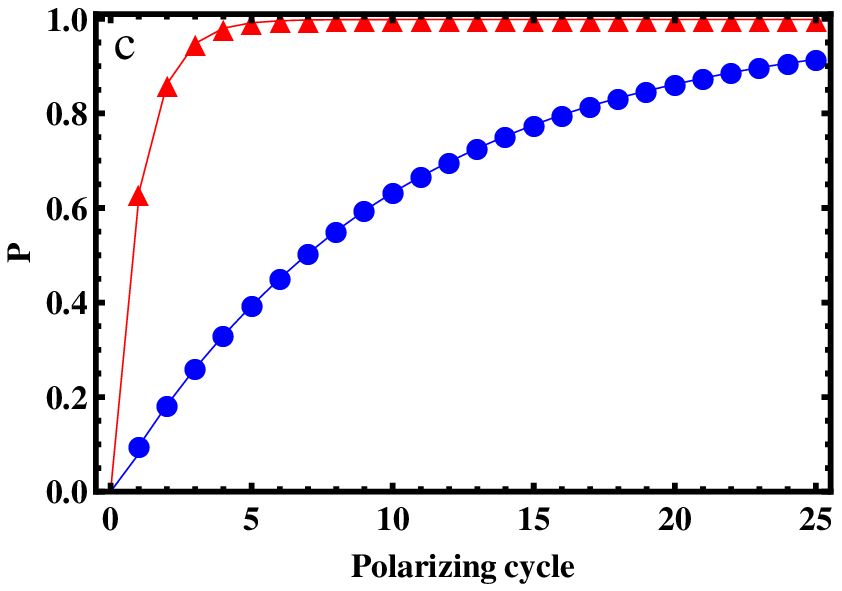}
\caption{Exact polarization dynamics with $B=0.36$ T, $v=(2\pi)6$ MHz$/\mu s$ and $\Delta t=10$ $\mu$s: a, The
coupling strength is given as $a_{x'}=(2\pi)0.6$ MHz, $\Omega_{eff}=(2\pi)3$ MHz (blue circles) and
$\Omega_{eff}=(2\pi)3.6$ MHz (red triangles). b, We have $\Omega_{eff}=(2\pi)3.5$ MHz, $a_{x'}=(2\pi)0.7$ MHz
(red triangles), $a_{x'}=(2\pi)0.5$ MHz (blue circles) and $a_{x'}=(2\pi)0.3$ MHz (green cubes). c, Exact polarization
dynamics with $v=(2\pi)0.8$ MHz$/\mu s$, $\Delta t=50$ $\mu$s, and $\Omega_{eff}=(2\pi)2.3$ MHz, $a_{x'}=(2\pi)0.3$ MHz
(red triangles) and $a_{x'}=(2\pi)0.1$ MHz (blue circles).}
\label{polarization}
\end{figure*}

So far we have assumed that the optical initial state $|0\rangle$ is well prepared. However, for
the relatively high magnetic fields that we are considering, the optical initial state is given
by $|0\rangle_{\theta}=\cos\theta|0\rangle + \frac{\sin\theta}{\sqrt{2}}
(e^{i\phi}|+1\rangle-e^{-i\phi}|-1\rangle)]$, (see eq. (\ref{optical})). For a perfect adiabatic sweep we obtain the average polarization in the
range $\theta\in[0^\circ,20^\circ]$ as
\begin{equation}
    \overline{P_{N_1}}=\frac{\int_{S}\Big(\cos^2\theta-\frac{\sin^2\theta}{2}\Big) dS}{S}.
\end{equation}
Here $S$ is the surface area of the two sectors within $\theta$ deviation and $dS$ is the
corresponding area element. The reachable polarization for nearby nuclear spins and the
average polarizations are $\overline{P_{N_1}}=0.91$ for $\theta\in[0^\circ,20^\circ]$.

\subsection{Polarizing cycles and effective ranges}

We can now summarize the polarizing cycles as follows (see also the black box of the
flow diagram Fig. \ref{flow1}): (i) Optical pumping initializes the NV spin and (ii)
a subsequent adiabatic transfer brings the electron spin to the $|-1\rangle$ state.
(iii) By using off-resonant driving and an quasi-adiabatic sweep across two Hartmann-Hahn
resonance points (ISE technique) polarization is transferred from the NV spin to the
nuclear spin. The steps (i)-(iii) are repeated as required. This method has two primary
advantages: (1) the impact of the random orientations is minimized, and NV spins in a
wide range of orientations are still accessible for polarization. (2) Using this method,
the coupling strength between the NV and nuclear spins is doubled because it involves
the double quantum transitions between the states $|-1\rangle$ and $|+1\rangle$.


In Fig. \ref{polarization} we simulate many cycles of this polarization protocol for a system of
one NV center and a single nuclear spin under the assumption that the NV center is initialized to
the state $|-1\rangle$ as an example. The initial density matrix of the unpolarized nuclear spin
is $\rho_0 = \mathbf{I}/2$, with $\mathbf{I}$ being the unit matrix. For the state $\rho_k$ of
the nuclear spin after the k-th iteration of the polarization protocol we have the following
relation
\begin{eqnarray}
    \label{evolution}
    \rho_{k+1} = tr_e[U_t(\rho_k\otimes|\chi_-\rangle\langle\chi_-|)U^{\dag}_t]
\end{eqnarray}
where $U_t=e^{-\int iH_{trans}dt}$ is time evolution operator, $H_{trans}$ is the Hamiltonian
eq. (\ref{Htrans}), we denote by $\Delta t$ the sweep time and by $tr_e$ the trace over the
electron spin. The polarizing P is defined as $P=\langle I_{z'}\rangle/\langle I_{z'}\rangle_0$
with $\langle I_{z'}\rangle$ denoting the expectation value of the nuclear spin and
$\langle I_{z'}\rangle_0$ the expectation value of the completely polarized state.

Finally, as we have discussed in the previous sections, the ISE requires an adiabatic slow passage.
For $\theta\in[0^\circ,20^\circ]$ we find $\delta(\theta)\in[0,(2\pi)45MHz]$ and in order to include
the entire range of detunings in the sweep we can estimate the required polarizing time. At a sweep
rate of $v=(2\pi)6$ MHz$/\mu s$ we find $\Delta t \sim10$ $\mu$s. In Fig. \ref{polarization}a, assuming
the perfect initialization of the NV spin in the state $|-1\rangle$, we take one nuclear spin with
the typical coupling $a_{x'}=(2\pi)0.6$ MHz as an example to demonstrate the polarization transfer
(the nuclear spin is about 0.5 nm from the NV spin and $a_{z'}=(2\pi)0.64$ MHz). Here the polarization
transfer after the first cycle, $P_1$, does not quite achieve the maximum possible value in Fig. \ref{pmax}
due to the existence of the phase shift $\Phi_{St}$, more precisely $P_1=0.58<P_{max}=0.7$ when
$\Omega_{eff}=(2\pi)3$ MHz, see both Fig. \ref{pmax}b and \ref{polarization}a. Simulations for different
hyperfine couplings are shown in Fig. \ref{polarization}b. Again the sweep rate $v=(2\pi)6$ MHz$/\mu s$
and polarization is transferred to the nuclear spins with different coupling to the NV spin, which agree
with Fig. \ref{pmax}. Additionally, slowing down the ISE sweep, as shown in Fig. \ref{polarization}c for
a sweep rate $v=(2\pi)0.8$ MHz$/\mu s$ we can polarize more distant nuclear spins with a coupling strength
$a_{x'}=(2\pi)0.1$ MHz, which corresponds to a distance of about 1 nm from the NV center. This coupling is
approaching the limitation of our polarization scheme, since the rotating-frame spin-lattice relaxation
time of NV spin in nanodiamonds is estimated to be about 100 $\mu$s which places a lower limit of the
duration of the full sweep~\cite{Laraoui}.


\begin{figure*}[t]
\center
\includegraphics[width=6.9in]{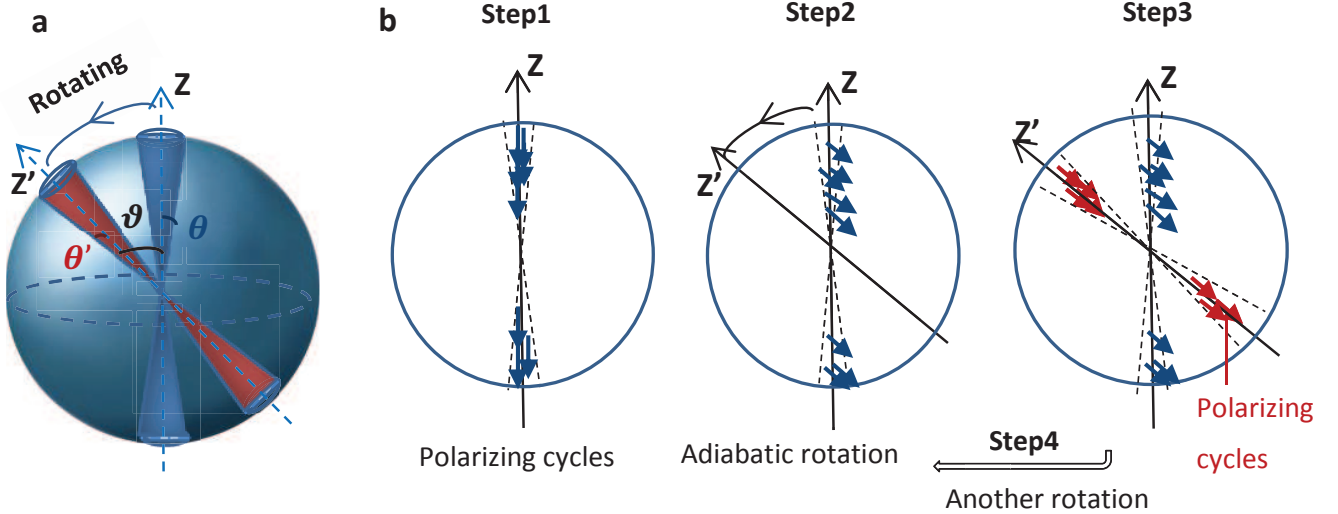}
\caption{a, Diagram of adiabatic rotating direction of the magnetic field. Blue spherical sectors present
$\theta$ degrees deviation tolerance for Z-direction of the magnetic field. Magnetic field is rotated by
angle $\vartheta$ to get red spherical sectors with the half-angle of the cone angle $\theta'$ to be involved.
b, Flow diagrams of polarizing steps of our scheme related to the adiabatic rotations of the magnetic field.
The small blue and red arrows are the nuclear spins nearby the different NV spins in different spherical
sectors. Four steps are given for our polarizing scheme. We use 2D plotting for simplicity, the sphere is
presented by a circle and the areas formed by the dashed lines denote the corresponding spherical sectors.
It is shown by the arrows that all the nuclear spins follows the rotated magnetic fields.}
\label{flow2}
\end{figure*}

\section{The benefit of adiabatic rotations}
\label{Adiabatic rotations for more polarizations}
The NV polarization in our scheme is limited by the adiabatic sweep range and efficiency of the
initialization of NV spins, enabling the polarization up to a maximum angle between NV center and
external magnetic field. As discussed above, our scheme certainly performs well for NV centers
whose orientation forms an angle of less than $20$ degrees from the externally applied magnetic
field (and may potentially work for somewhat larger angles). It is however desirable to extend
the polarization to a larger fraction of the nanodiamond ensemble, ideally covering the entire
solid angle. One method for achieving this applies an adiabatic change of the orientation of the
magnetic field relative to the NV center, either by rotation of the magnetic field or by rotation
of the nanodiamond. For example, following the magnetic field rotation to the $Z'$-direction, another
set of NV centers with orientations in a solid angle (shown in red in Fig. \ref{flow2}) around
the new magnetic field direction will participate in the polarization dynamics while the NV centers
in the original orientations (the blue cone in Fig. \ref{flow2}) will be inactive. Hence previously
polarized nuclear spins will not affected by the polarization sequences, but these nuclear spins
may use this "idle" time to spread their polarization by spin diffusion across larger volumes
until further rotations subject them to another polarization cycle.

More specifically, let us consider the rotation of the magnetic field as an example. The Hamiltonian
that is describing the rotation of the magnetic field from the $Z$-direction to the $Z'$-direction
is given by
\begin{eqnarray*}
    H_{B} &=& \gamma_NB_{Z}[\cos(\dot{\vartheta}t)I_{Z}+\sin(\dot{\vartheta}t) I_{X}]
    = \gamma_NB_{Z}I_{\vartheta(t)}
\end{eqnarray*}
where $\vartheta(t)=\dot{\vartheta}t$ is the angle between the rotated magnetic field and its
initial direction and $\gamma_NB_{Z}=(2\pi)4$ MHz.

Similar to the discussion in the initialization scheme, the well-known adiabatic condition
requires $\dot{\vartheta}\ll2\gamma_NB_{Z}$. For a rotation by $\vartheta=180^\circ$ this
leads to the condition $t_r\gg0.078$ $\mu$s for the minimal rotation time $t_r$. This implies
that $t_r\cong 0.78$ $\mu$s will ensure adiabaticity and minimal perturbation of nuclear
spin polarization for a rotation to an arbitrary orientation. This rotation time is much
shorter than even a single polarization cycle. Hence, after running a few polarization cycles
that achieve a high net polarization of nuclear spins for NV centers oriented along the $Z$
direction, we can adiabatically rotate the magnetic field to another direction. The polarization
of the nuclear spins will then be achieved for this new direction, as shown in Fig. \ref{flow2}.
Note though that the rapid rotation of a strong magnetic field is challenging. Hence one
needs to resort either to a mechanical rotation of the sample as a whole or, as explained
in the next section, make use of the natural Brownian rotation of nanodiamonds in solution.

In addition to the rotation of the externally applied magnetic field for nanodiamond powder, another option
consists of the use of random Brownian rotations of nanodiamonds in a solution.
%
If the Brownian rotation is sufficiently slow to satisfy our adiabatic rotation condition, the polarized
nuclear spins will maintain their polarization in the direction of the externally applied magnetic field.
The time scale of Brownian rotations is determined by~\cite{Frenkel}
\begin{equation}
\tau_B=\frac{3V_H\eta}{kT},
\end{equation}
where $k$ is the Boltzmann constant, $T$ is the temperature, $V_H$ is the hydrodynamic volume of the particle
and $\eta$ the viscosity of the surrounding carrier liquid. Note that the hydrodynamic volume is an effective
volume that includes both the true particle volume and the volume of a fluid that is displaced when the particle
rotates due to particle-fluid interaction. A typical average hydrodynamic diameter for nanodiamonds of $30$ nm
in water gives a Brownian relaxation time of $\tau_B\simeq9$ $\mu$s, which is comparable with our polarizing
cycle time. One can increase the Brownian relaxation time by using nanodiamonds with a larger diameter, a more
viscous fluid or lower temperatures. Additionally, surface coating can increase the average hydrodynamic diameter,
i.e., after adsorption of ferritin the diameter of nanodiamond was determined to be $85$ nm~\cite{Ermakova}.

Here we assume that the hydrodynamic diameter of nanodiamond is $85$ nm, and the Brownian relaxation time is
$\tau_B\simeq205$ $\mu$s. On average after a time $\tau_B$, the nanodiamonds will have rotated to another random
direction, and different NV centers will now participate in the polarization dynamics of nearby nuclear spins
while other NV centers will be inactive as they are effectively decoupled from their neighboring nuclear spin
due to the absence of a Hartmann-Hahn condition. When an NV center "leaves" the coherent exchange range via
a random rotation, the polarization of the polarized nuclear spins will still be transferred to the other
nuclear spins by spin diffusion. Considering maximum of 20 degrees deviation, 6\% percents of the NV spins
are involved at each given time, and each NV will be in this range after 16 random rotations on average.

\begin{figure*}[t]
\center
\includegraphics[width=2.3in]{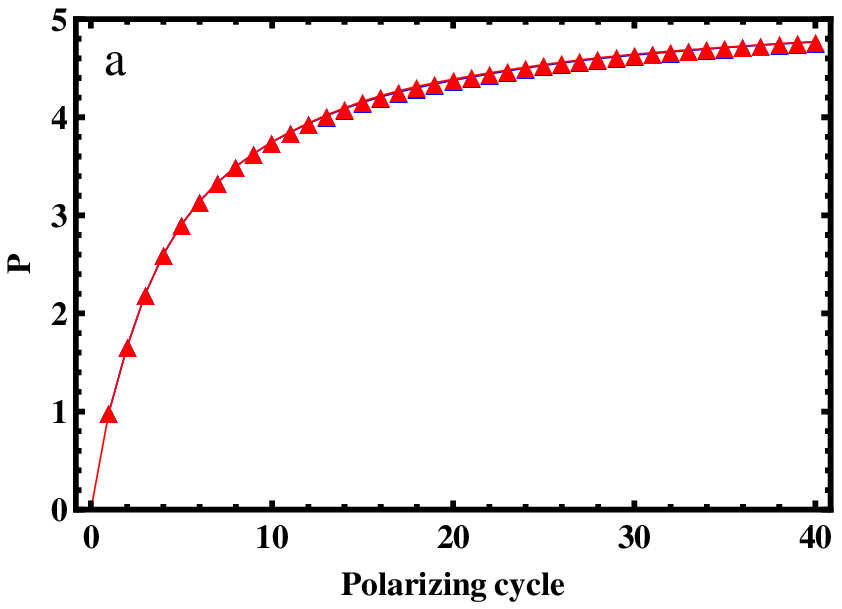}
\includegraphics[width=2.3in]{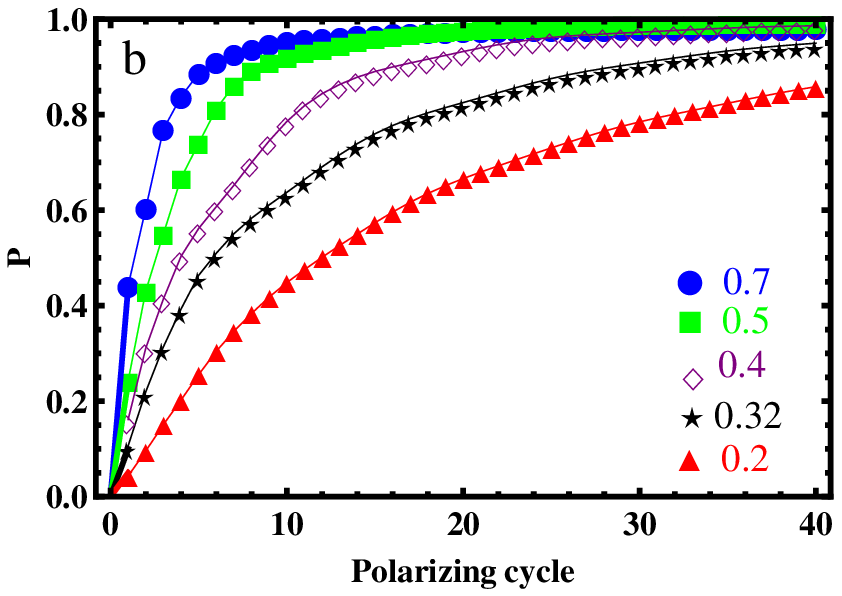}
\includegraphics[width=2.3in]{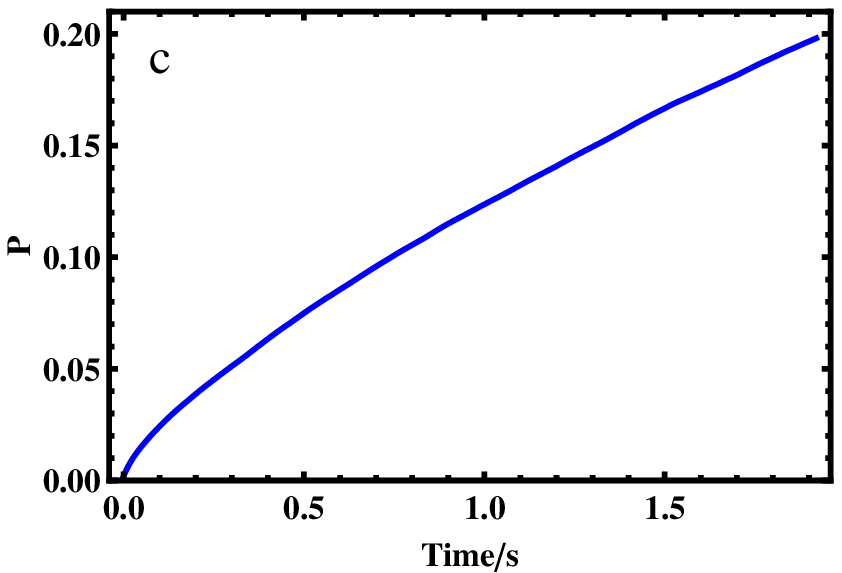}
\caption{The polarization buildup for five nuclear spins for $\Omega_{eff}=3.23$ MHz. The nuclear
spins are strongly interacting with the NV center spin, $a_{x'_{1}}=(2\pi)0.7$ MHz, $a_{x'_{2}}=(2\pi)0.5$ MHz,
$a_{x'_{3}}=(2\pi)0.4$ MHz, $a_{x'_{4}}=0.32$ MHz, and $a_{x'_{5}}=(2\pi)0.2$ MHz: a, The blue and red triangles
denote the presence and absence of the interaction among the nuclear spins. Internuclear interactions do not lead
to significant changes in the polarization dynamics because the dipole-dipole interaction of nuclear spins is much
smaller than the electron-nuclear coupling. b, The polarization dynamics of the individual nuclear spins in the
set of five spins represented by different markers within different colors. c, Numerical simulation of the polarization
buildup in a large number of polarization cycles in a system consisting of an NV spin and around $548$ $^{13}C$ nuclear
spins in a volume of $50000$ lattice sites by using Brownian rotation. The internuclear dynamics is treated within
the bosonic approximation and we simulate a polarization buildup time of $2$ second. The system benefits from
polarization diffusion between nuclear spins resulting in a high level of polarization.}
\label{multi}
\end{figure*}

\section{Internuclear interaction and spin diffusion}
\label{Effect of nuclear dipolar interaction}
So far, we have considered polarization of an individual nuclear spin and have therefore ignored the effects
of inter-nuclear coupling on both the polarization cycle and on the nuclear spin polarization diffusion towards
more distant nuclear spins that are not directly interacting with the NV center. This section discusses these
effects of nuclear dipolar interactions. Considering our off-resonant driving in the dressed state basis and
the suitable rotating frame, the effective Hamiltonian of the whole system consisting of an NV spin and its
nearby nuclear spins can be written as
\begin{eqnarray}
    \label{total}
    H_{tot}\approx H_{NV}+H_{hf}+H_{d-d}+H_{Nu}.
\end{eqnarray}
Here $H_{NV}$ denotes the NV center effective energy Hamiltonian, $H_{hf}$ the effective electron-nuclear hyperfine
interaction, $H_{d-d}$ the dipole-dipole coupling among nuclear spins and $H_{Nu}$ is the magnetic field splitting
of nuclear spins
\begin{eqnarray}
    H_{NV} &=& 2\Delta(t) \sigma_x+\Omega_{eff}\sigma_{z}, \\
    H_{hf} &\simeq& \sum_{i}\sigma_{x}(2a_{x'_{i}}I^{x'}_{i}+2a_{z'_{i}}I^{z'}_{i}) \\
    H_{d-d} &\simeq& \sum_{ij,j>i}d_{ij}\Big[\vec{I}_i\vec{I}_j-3(\vec{I}_i\cdot\vec{e}_{r_{ij}})(\vec{I}_j\cdot\vec{e}_{r_{ij}})\Big]\\
    H_{Nu} &=& \sum_{i}\gamma_{n}BI^{z'}_{i},
\end{eqnarray}
with $a_{x'_{i}}$ and $a_{z'_{i}}$ denoting the hyperfine coupling strength between the $i$th nuclear spin and the
electron spin. $d_{ij}=\frac{\mu_0}{4\pi}\frac{\gamma_n\gamma_n}{r_{ij}^3}$ is the coupling strength between the $i$
and $j$ nuclear spins.  with $r_{ij}=|\vec{r}_{ij}|$ denoting the distance between the two nuclear spin and
$\vec{e}_{r_{ij}} = \vec{r}_{ij}/r_{ij}$. The $i$ and $j$ indices are
summed over all nuclei.

In order to estimate the effect of internuclear interactions on the polarization transfer from the NV center to the
nuclei in the multiple nuclear spins case we consider 5 nuclear spins and one electron spin, including hyperfine
and nuclear dipolar couplings. After each polarization cycle, the electron spin is re-initialized in the state
$|-1\rangle$. The concatenated evolution of the nuclear spin density matrix is determined by an equation analogous
to eq. (\ref{evolution}).
The initial density matrix of the five unpolarized nuclear spins can be written as $\rho_0=\mathbf{I}/2^5$, with
$\mathbf{I}$ being a unit matrix of dimension $2^5$, and the time evolution is given by the operator $U_t=e^{-i\int H_{tot}dt}$,
with $H_{tot}$ denoting the total Hamiltonian eq. (\ref{total}). The polarization is defined as in section II,
$P_k=\langle I^k_{z'}\rangle/\langle I^k_{z'}\rangle_0$ in which the subscript $k$ presents different nuclear
spins and $P=\frac{1}{5}\sum_k^5P_k$. We assume in these simulations only nearest neighbour inter-nuclear spin
coupling with rates $d_{12}=d_{23}=d_{34}=d_{45}=2$ kHz.

The simulation results are presented in Fig. \ref{multi}a and b. We consider the case that the NV is strongly
coupled to all five $^{13}C$ spins, $a_{x'_{1}}=(2\pi)0.7$ MHz, $a_{x'_{2}}=(2\pi)0.5$ MHz, $a_{x'_{3}}=(2\pi)0.4$ MHz,
$a_{x'_{4}}=0.32$ MHz, and $a_{x'_{5}}=(2\pi)0.2$ MHz. The polarization is built up for all the five nuclear spins
with different coupling to the NV spin, which matches our theory. In this case, the coupling between the nuclear
spins is too small to affect the evolution of the polarization dynamical built-up (Fig. \ref{multi}a). Due to small
interaction among the nuclear spins, there is almost no difference between polarization in the presence or absence
of the interactions among the nuclear spins, which means our scheme works in the frozen core.

Now we would like to exemplify the benefits of internuclear interactions which leads to
nuclear spin diffusion and a considerable extension of the range of polarization.
Suppose a typical configuration of nuclear spins distributed around the NV. To estimate
the polarization efficiencies in a large system, we adopted the simple spin temperature
approximation, namely neglecting the nuclear spin coherence~\cite{Cirac}, which provides
a good estimation of polarization transfer efficiency when the sweep time step is sufficiently
small. This approximation yields independent rate equations for each individual nuclear
spin and should give a conservative estimate for the achieved polarization. For the $\Delta t$
sweep times described above, this approximation serves as a lower bound, as the achieved
polarization in one cycle with our scheme is higher than the polarization achieved during
the same time with small sweep steps.

For the diffusion step, the interaction between the nuclear spins is taken into account,
where we use the Gaussian approximation for the nuclear spins. As is typical for paramagnetic
centers, the NV center strongly affects the spin diffusion in nearby nuclear spins (the
"frozen core") due to the energy mismatch caused by the $S_zI_z$ hyperfine term. However,
when the NV center is in the $|m_s = 0\rangle$ state, this energy mismatch is suppressed
and diffusion to external nuclear spins is allowed. To give a conservative estimate, we
assume that no diffusion in the frozen core takes place when the NV center is in the
$|m_s = \pm 1\rangle$ states. For this purpose, we simulate the NV center as a classical
spin with the probability of being in each eigenstate given by the optical initialization
corresponding to the NV orientation.

As the nanodiamonds are in a solution, we assume that the differences in polarization transfer
for the different orientations of the NV spins within the 20-degree deviation range can be
neglected. Consider a random configuration of nuclear spins, with $548$ nuclear spins in the
vicinity of an NV center (lattice of 50,000 nuclei per NV center) whose orientation is within
the 20-degree deviation range. The net polarization is built up by using Brownian rotations
as shown in Fig. \ref{multi}c. There are two steps in the simulation. First, we assume that
the orientation of the NV center satisfies $\theta\in[0^\circ,20^\circ]$ and we consider
$\Omega_{eff}=(2\pi)2.3$ MHz, sweep rate of $v=(2\pi)0.8$ MHz$/\mu s$ and $\Delta t=70$
$\mu$s. The polarization transfer from the NV center to the neighboring nuclear spins is
achieved by using our combination of off-resonantly driven double quantum transition and
the ISE technique. The polarizing cycles are repeated during the Brownian relaxation time
$\tau_B=205$ $\mu$s. Secondly, after the Brownian relaxation time, the nanodiamond is randomly
rotated to another direction. As we discussed in the Sec. \ref{Adiabatic rotations for more polarizations},
it takes roughly $3.5$ ms for the nanodiamond to rotate back to our maximal deviation range for
polarization, and nuclear spin diffusion dominates during this time period (taking into account
the NV effect on diffusion in the frozen core). In this step more distant nuclear spins can become
polarized by diffusion. These two steps occur sequentially over a $2$ second time range. In line with
previous sections, the polarization is defined as $=\langle I_{z'}\rangle/\langle I_{z'}\rangle_0$,
where $I_{z'}=\sum_i^MI^{z'}_i$, $M$ is the number of the involved nuclear spins. Interestingly,
the slope of polarization is close to linear. This demonstrates that in the long diffusion time,
where the nanodiamond is not involved in the polarization cycles, we do not see a bottleneck
caused by slow diffusion and significant polarization is transferred from the NV spin to
the nuclear spins within each step 1. The final net polarization reaches $0.2$, equivalent to
about $110$ $^{13}C$ fully polarized spins are polarized within $2$ second in our scheme.

\section{A large angle polarization scheme}
\label{A interesting range of the NV orientations}
So far we have concentrated so far on a range of angles $\theta\in[0^\circ,20^\circ]$ between NV center and the
external magnetic field. Note however, that this is not the only range in which efficient polarization transfer
can be achieved. Indeed, we have another interesting range of the NV orientations, concentrated around the direction
perpendicular to the magnetic field, i.e., $\theta\in[70^\circ,110^\circ]$, which involves 34\% the NV spins.
Here the condition $D(\theta)<0$ leads to the Hamiltonian $H_{-}$ in eq. (\ref{H plus}). Following steps that
are analogous to those leading up to Hamiltonian eq. (\ref{Htrans}) we find for the subspace $\{|\chi_-, \downarrow\rangle, |\chi_+, \uparrow\rangle\}$
\begin{equation}
    H_{matrix}	=	
    \left(\begin{array}{cc}
        \omega_{eff}-\frac{\gamma_{n}B}{2} & \frac{a_{x'}\sin\varphi}{2}\\
        \frac{a_{x'}\sin\varphi}{2} & -\omega_{eff}+\frac{\gamma_{n}B}{2}
    \end{array}\right).
\end{equation}
Suppose now that the NV center is initialized in state $|-1\rangle$ by using the off-resonant driving. Then we
make use of the ISE technique to sweep adiabatically across the two possible Hartmann-Hahn resonance points of
the dressed NV spin and nuclear spin pair. As a result the the population is transferred between $|\chi_-,\downarrow\rangle$
and $|\chi_+,\uparrow\rangle$, with the other states being unaffected. As before this allows the nuclear spins to
be polarized but now in a direction opposite to the case $\theta\in[0^\circ,20^\circ]$. In the following we
discuss the details of this case.

\subsection{Initial polarization of the NV spins}
We begin with a discussion of the initialization of the NV center spins. As explained in previous sections
optical pumping of the NV spins results in the $|m_s=0\rangle$ state in the NV frame, that is
\begin{displaymath}
    |0\rangle_{\theta} = \cos\theta|0\rangle +\frac{\sin\theta}{\sqrt{2}}(e^{i\phi}|+1\rangle-e^{-i\phi}|-1\rangle)
\end{displaymath}
where the states $|0\rangle,|\pm 1\rangle$ are defined in the laboratory frame. For $\theta\in[70^\circ,110^\circ]$
the population in state $|0\rangle$ is very small and the initial state is well approximated by
$\frac{1}{\sqrt{2}}(e^{i\phi}|+1\rangle-e^{-i\phi}|-1\rangle)$ which, in the relevant
subspace spanned by $\{|+1\rangle,|-1\rangle\}$, is unpolarized.

We address this problem by means of an adiabatic sweep. For concreteness, consider $\Omega=(2\pi)20$ MHz and
an adiabatic sweep from $\Delta_{WM}(t_i)=(2\pi)320$ MHz to $\Delta_{WM}(t_f) <-(2\pi)320$ MHz. For a sweep range
$|\omega_1(t_f)-\omega_1(t_i)|\simeq (2\pi)640$ MHz, the time for an adiabatic sweep is of the order of $0.4$ $\mu$s.
Such a sweep induces an adiabatic population transfer from state $|-1\rangle$ to state $|0\rangle$ so that the
state of the NV center evolves to
\begin{displaymath}
    |0\rangle_{\theta}{\longrightarrow}\cos\theta|-1\rangle + \frac{\sin\theta}{\sqrt{2}} (e^{i\phi}|+1\rangle - e^{-i\phi}|0\rangle)
\end{displaymath}
where we ignore any dynamical and geometric phase that has been accumulated in the sweep. It is now crucial
to note that the state $|0\rangle$ does not contribute to the polarization dynamics as it does not take part
in the far detuned dynamics that is induced by the applied microwave fields. Hence the relevant quantity is
the polarization in the subspace spanned by the states $\{|+1\rangle, |-1\rangle\}$. Normalized by the population
that is found in this subspace we find
%
%
%
\begin{equation}
    \overline{P^r_{N_2}}=\frac{\int_{S}N_r\Big(\frac{\sin^2\theta}{2}-\cos^2\theta\Big)  dS}{S},
\end{equation}
where $N_r=1/(\frac{\sin^2\theta}{2}+\cos^2\theta)$ is the normalized coefficient and $\overline{P^r_{N_2}}\sim0.85$.

\subsection{The sweep range for ISE}

Owing to the different behaviour of the energy shifts $D(\theta)$ and $\delta(\theta)$ we need to change the sweep
range in the ISE. In order to achieve optimum polarization transfer, we choose $\Omega=(2\pi)40.5$ MHz, which results
in a range of effective Rabi frequency $\Omega_{eff}\in[(2\pi)2.3MHz, (2\pi)3.6MHz]$. The adiabatic slow passage
needs to cross the range $\Delta\in[-(2\pi)21MHz,(2\pi)21MHz]$. For a sweep rate of $v=(2\pi)6$ MHz$/\mu s$ we estimate
the required polarizing time as $\Delta t>7$ $\mu$s. Apart from these parameter changes, the principles underlying our
discussion of the polarization transfer in Section II are still valid.

If we assume that the NV spin is initial polarized perfectly, we have $P_{max}>0.2$ in a coupling range $a_{x'}\in[0.4,0.9]$
for $|\Omega_{eff}|=(2\pi)2.27$ MHz in Fig. \ref{pmax}c, compared to $a_{x'}\in[0.18,0.9]$ for $|\Omega_{eff}|=(2\pi)3.6$ MHz in Fig. \ref{pmax}a.
We can say that our off-resonant driving and ISE technique allow polarization transfer for all the NV orientations
within $\theta\in[70^\circ,110^\circ]$. Consider the average polarization of the NV spins in this case is about 0.5
and twice effective Rabi frequency range here, applying a singe sweep speed becomes not very efficient for all the
NV orientations, especially for the nuclear spins with weak couplings to the NV spin. As we discussed, we have another
choice, slow down the ISE sweep can involve the nuclear spins with weaker coupling to the NV spin, i.e., as shown
in Fig. \ref{polarization}c, we can use another a speed rate $v=(2\pi)0.8$ MHz$/\mu$s ($\Delta t\sim50$ $\mu$s) to
make it possible to polarized the $^{13}C$ spins with coupling strength $a_{x'}=(2\pi)0.1$ MHz.

In conclusion, off-resonant driving and the ISE technique allow for polarization transfer between the NV spin and
its surrounding nuclear spins, when the orientation of the NV spin matches $\theta\in[70^\circ,110^\circ]$ but we
need to sacrifice half the population to provide a high initial polarization of the dressed NV spins. On the other
hand the fact that around $34\%$ of the NV centers participate in the polarization dynamics even without magnetic
field rotation is an attractive feature.

\section{Depolarizing effect}
\label{Depolarizing effect}
Our approach to increase the fraction of NV center that participate in the polarization sequences by means
of adiabatic rotations may potentially have an undesired side effect as nuclear spins that have been polarized
for one magnetic field orientation may be depolarized again when transferred to a different magnetic field
orientation. Consider for example the case shown in Fig. \ref{flow2} where, after polarization transfer is
achieved for the blue spherical sectors, the magnetic field is rotated from $Z$-direction to different direction
$Z'$. If the NV spins in the blue sectors have not been polarized in advance and our combination of off-resonant
driving and ISE achieves a Hartmann-Hahn condition at some instant, then the depolarized state of the NV center
will be transferred to nuclear spins, whose polarization reduces as a result. In the protocol for the case
$\theta\in[0^\circ,20^\circ]$ this depolarization is negligible due to the rapid increase of the detuning
$\delta(\theta)$ outside of this range. Only NV centers in the range $\theta\in[20^\circ,25^\circ]$ may still
experience undesired resonances. As these NV centers are still well initialized by means of optical pumping
depolarization is negligible.

The situation is different and indeed more complex for the second protocol that covers the range
$\theta\in[70^\circ,90^\circ]$. NV centers with orientations in the range $\theta\in[35^\circ,43^\circ]$
are difficult to be polarized to the required state. Unfortunately, they have almost the same energy
distributions $|D(\theta)|$ and $\delta(\theta)$ as the NV centers in the range $\theta\in[70^\circ,110^\circ]$,
see Fig. \ref{distribution}. Our off-resonant driving and ISE technique scheme can lead to Hartmann-Hahn resonances
at some instant which then cause polarization exchanges between the unpolarized NV centers and the polarized
nuclear spins.
Thus, magnetic field rotations could lead to depolarization dynamics in the previously polarized nuclear
spins for the protocol that is adapted to the case $\theta\in[70^\circ,110^\circ]$.

\begin{figure}
\center
\includegraphics[width=2.8in]{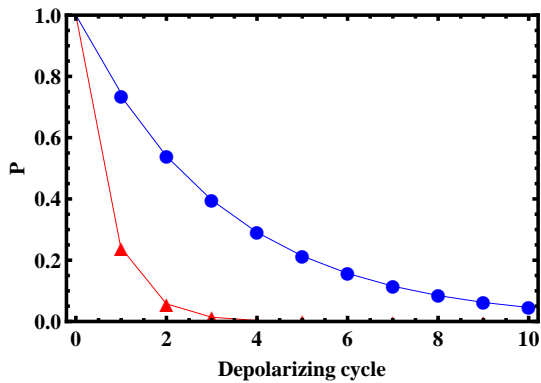}
\caption{The depolarization of the previously polarized nuclear spin with $B=0.36$ T, $\Omega_{eff}=(2\pi)3.5$ MHz, $v=(2\pi)6$ MHz$/\mu s$ and $\Delta t=10$ $\mu$s. The coupling strength is given as $a_{x'}=(2\pi)0.3$ MHz, (blue circles) and $a_{x'}=(2\pi)0.6$ MHz (red triangles). }
\label{depolarization}
\end{figure}

We consider a simple system consisting of one NV and one nuclear spin as an example for studying the depolarizing
effect of this case. The depolarizing cycle time is as the same as our polarizing cycle time, since every initialization
process is implemented on all the NV spins. Consider a nuclear spin that is initially polarized and in contact
with an unpolarized electron spin $\rho_{e0}=\mathbf{I}/2$. The evolution of the nuclear spin density matrix is
then given by
\begin{eqnarray}
    \rho\rightarrow \cdots U'_ttr_e[U'_t(|\downarrow\rangle\langle\downarrow|\otimes\rho_{e0})U'^{\dag}_t]\otimes\rho_{e0}U'^{\dag}_t\cdots,
 \end{eqnarray}
where $U'_t=e^{-i\int H^{\delta'}_{matrix}dt}$ is time evolution operator. $H^{\delta'}_{matrix}$ is given by
eq. (\ref{Htrans}) with the choice $\Omega_{eff}=(2\pi)3.5$ MHz and $\delta'$ denoting that a magnetic field
in $Z'$-direction is being considered. The polarization P of the nuclear spin is defined as in Sec. II. As
shown in Fig. \ref{depolarization}, the nuclear spin is depolarized very rapidly.
The depolarization range $\theta'\in[35^\circ,43^\circ]$ includes $9\%$ of the NV orientations while the
range in which the polarization is active includes $34\%$ of the NV centers. Hence, although adiabatic
rotation of the magnetic field brings an unexpected depolarizing effect, it is nevertheless quite attractive
as the fraction of NV spins that are contributing to the polarization transfer even without magnetic field rotation
strongly outweighs the small fraction of $9\%$ of the NV centers that suffer depolarization.

\section{Discussion}
\label{Discussion}

To summarize, our polarization scheme for ensembles of nanodiamonds (see Fig. \ref{flow1}
and Fig. \ref{flow2}):\\
Step 1: A moderate to strong magnetic field is applied along $Z$-direction (of order $B=0.36T$). A
polarization cycle consists of optically pumping and initialization of the NV center spins, and the
subsequent application of off-resonant microwave driving and the integrated solid effect to realise
a robust adiabatic sweep across Hartmann-Hahn resonance points. This achieves the coherent polarization
transfer from the NV center spins to the nuclear spins with a range of orientations of the NV center
relative to the magnetic field. This polarization cycle is repeated several times. \\
Step 2: Adiabatic rotations (of the magnetic field or of the nanodiamonds) change the magnetic field
orientation relative to the NV centers to a new $Z'$-direction, while the nuclear spins follow the
direction of the magnetic field.\\
Step 3: Repeat step 1 and step 2.\\

Apart from control imperfections, the achievable $^{13}C$-bath spin polarization will be limited
by the spin-lattice relaxation time $T_1$ of the nuclear spins. In recent experiments, by using
electron paramagnetic resonance (EPR), solid-state NMR, and DNP techniques, a variety of nanodiamond
samples, of varying manufacturing processes and particle sizes, were examined~\cite{Casabianca}.
Several types of nanodiamonds were identified that exhibit very long nuclear spin lifetimes of
several minutes. Additionally, this polarization transfer rests on NV spins having a sufficiently
long rotating-frame spin-lattice relaxation time $T_{1\rho}$ in nanodiamonds during the adiabatic
passages that are required for the implementation of ISE. Using a strong MW field driving,
$T_{1\rho}$ can be estimate as 100 $\mu$s~\cite{Laraoui,Belthangady}, which is much longer than
our sweep time in one polarization cycle. Assuming a polarization cycle duration of $\Delta t=$10 $\mu$s,
when rotating the magnetic field, $17$ rotations of the magnetic field are required to involve
all the NV spins. For a nuclear spin lifetime of more than 2 minutes, this would allow the execution
of around $10^5$ polarization cycles for each NV spin involved before the nuclear spins start to
relax. For nanodiamonds in a solution, each nanodiamond is involved in the polarization cycles
roughly once every $17$ random Brownian rotations. Assuming $\tau_B\simeq205$ $\mu$s, with a pure
nuclear diffusion period of roughly $3$ms after every Brownian relaxation time $\tau_B$, $2$ minutes
lifetime of the nuclear spin would allow in excess of $10^5$ polarization cycles for every NV center.

We can now obtain a rough estimate of the total amount of the net polarization that could
be achieved under realistic experimental conditions. Consider a $1$ mm$^3$ nanodiamonds
powder, dissolved in a solution. The nanodiamonds are assumed to have a diameter of $85$ nm,
with an NV center concentration of $C=2\times10^{18}$ $cm^{-3}$ (see~\cite{Su2013} and
references therein for achievable NV concentrations). In the limit of a uniform
spatial distribution, this gives about $642$ NV spins per nanodiamond, with the total amount
of NV spins given by $8.35\times10^{15}$ NV spins. Consider natural abundance of the $^{13}C$
spins (1.1\%), the total nuclear spins in this powder of nanodiamond is approximately
$8.18\times10^{18}$~\cite{Number}. Note that $^{13}C$ enriched nanodiamonds will enable
polarization of more nuclear spins with each NV center and hence a larger overall level
of polarization. Taking account of the high degree of polarization, in excess of $0.3$ as
demonstrated in Sec. \ref{Effect of nuclear dipolar interaction}, we find that a total
amount of polarization equivalent to $1.6\times10^{18}$ nuclear spins should be achievable
in such a sample within 2 seconds.



\section{Conclusion}
\label{Conclusions}
In conclusion, in the high magnetic field limit, we propose a scheme achieving macroscopic levels of
$^{13}C$ nuclear spin polarization in randomly oriented ensembles of nanodiamonds, realised as powder
or solutions. To address the lack of a common quantization axis for the NV centers, an off-resonant
microwave drive realises a resonant double quantum transition and this together with the integrated
solid effect enable microwave coupling and control of NV spins whose orientation deviates from the
external magnetic field by less than 20 degrees. Matching the effective Rabi frequency of the
off-resonant double quantum transition of the NV centers to the nuclear spin Larmor frequency, enables
near resonant coupling between NV electron spins and the nuclear spins, and the transfer of polarization
from the initialized NV spins to the neighbouring $^{13}C$ nuclear spins. Additionally, the effect of
nuclear dipole-dipole interactions allows the weakly coupled nuclear spins to be polarized by spin
diffusion. Adiabatic rotation (of the magnetic field or sample) can then extend the polarizing scheme
to more NV spins, achieving a net polarization of all  nuclear spins via our scheme. By using our
polarization scheme, room-temperature, optical based polarization of the nuclear spins is made possible.

These results introduce several exciting opportunities for the application hyperpolarized nanodiamonds,
especially in the biomedical sciences. Diamond nanoparticles are biocompatible, exhibit no \textit{in vivo}
toxicity~\cite{Zhu}, and can be attached to a wide range of specific proteins, peptides or antibodies
\cite{Mochalin} while maintaining long T1 times. Thus, hyperpolarized nanodiamonds present an exciting
platform as MRI probes for molecular imaging.


\section*{Appendix: effective Hamiltonian of an NV center at the high magnetic field limit}
\setcounter{figure}{0}
\renewcommand{\thefigure}{A\arabic{figure}}
\begin{figure}
\center
\includegraphics[width=2.8in]{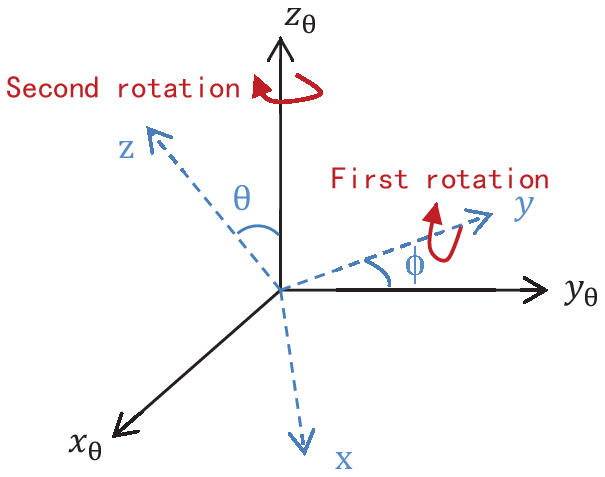}
\caption{Two related rotations of the coordinate system are used to map from the natural orientation
$z_{\theta}$-axis of the NV spin to z-axis which is defined by the high magnetic field. }
\label{figA1}
\end{figure}

In our work we assume that in the laboratory frame of reference a strong magnetic field is applied
along the $z$-direction, $\gamma_{e}B_{z}\gg D$. In the following we derive the representation of
the NV center Hamiltonian eq. \ref{hamNV} in the laboratory frame of coordinates. To this end we
need to transform the zero-field splitting tensor, which is diagonal in the natural coordinate system
defined by the NV center axes, to the laboratory frame of reference. We assume that the angle between
the magnetic field, i.e. the $z$-direction, and the NV-axis, the $z_{\theta}$-direction is given by
$\theta$ and that the angle between the laboratory frame $y$-direction and the NV center $y_{\theta}$
directions is given by $\phi$ in the x-y plane, as shown in Fig. \ref{figA1}. The spin operators in
the NV system of coordinates, $S_{x_{\theta}}$, $S_{y_{\theta}}$ and $S_{z_{\theta}}$, are then
related to the spin operators $S_{x}$, $S_{y}$ and $S_{z}$ in the laboratory coordinates by
\begin{eqnarray*}
    S_{x_{\theta}} &=& \sin\theta S_{z} + \cos\theta(\cos\phi S_{x} - \sin\phi S_{y}),\\
    S_{y_{\theta}} &=& \cos\phi S_{y} + \sin\phi S_{x},\\
    S_{z_{\theta}} &=& \cos\theta S_{z} - \sin\theta (\cos\phi S_{x} - \sin\phi S_{y}).
\end{eqnarray*}
The Hamiltonian of the electron spin
\begin{eqnarray} \nonumber
    H^e_N &=& (-\frac{1}{3}D+E)S_{x_{\theta}}\cdot S_{x_{\theta}} + (-\frac{1}{3}D-E)S_{y_{\theta}}\cdot S_{y_{\theta}}\\
    \nonumber
    &&+(\frac{2}{3}D)S_{z_{\theta}}\cdot  S_{z_{\theta}}+\gamma_{e}BS_z,
\end{eqnarray}
can then be rewritten in the basis defined by the eigenstates of the $S_z$ operator of the electron spin,
$\{|+1\rangle,|0\rangle,|-1\rangle\}$ as
\begin{widetext}
    \begin{eqnarray}
    \nonumber
    H''_{NV} =
    \left(\begin{array}{ccc}
        D(\theta)+\gamma_{e}B_{z} & -G_1 & G_2\\
        -G_1^* & 0 & G_1\\
        G_2^* & G_1^* & D(\theta)-\gamma_{e}B_{z}
    \end{array}\right)
\end{eqnarray}
 \end{widetext}
with
\begin{eqnarray}
    \nonumber D(\theta)&=&\frac{D(1+3\cos(2\theta))+3E(1-\cos(2\theta))}{4},\\ \nonumber
    G_1&=&\frac{(D-E)\sin\theta\cos\theta e^{i\phi}}{\sqrt{2}} ,\\ \nonumber
    G_2&=& \frac{[D+3E+(E-D)\cos2\theta ]e^{2i\phi}}{4}.
\end{eqnarray}

There exist two uncontrollable components, the angles $\theta$ and $\phi$ as well as the local
strain $E$, which modifies the spin level structure.

As we know that a high magnetic field suppresses the effect of the off-diagonal terms in $H''_{NV}$.
Assume that the Hamiltonian is of the form $H''_{NV}= H''_{NV0}+\epsilon V$, where $H''_{NV0}$ is
the diagonal part of the Hamiltonian and $\epsilon V$ the off-diagonal part. In our case the magnetic
field is sufficiently large for us to assume that the off-diagonal part is a weak perturbation. By
using the Schrieffer-Wolff transformation in condensed matter, the second order corrections due to
the off-diagonal terms $\epsilon V$ can be obtained as
\begin{widetext}
 \begin{eqnarray}
    \nonumber
    \langle\alpha|H''_{M}|\beta\rangle=\frac{\epsilon^2}{2}(\frac{\langle\alpha|V|i\rangle\langle i|V|\beta\rangle}{E_\alpha-E_i}-\frac{\langle\alpha|V|i\rangle\langle i|V|\beta\rangle}{E_i-E_{\beta}}),
\end{eqnarray}
\end{widetext}
in which $H''_M$ is defined as the corrected Hamiltonian. By simple calculation, we find
 \begin{eqnarray}
    \nonumber
    \langle+1|H''_{M}|+1\rangle=\frac{|G_1|^2}{\gamma_e B+D(\theta)}+\frac{|G_2|^2}{2\gamma_e B},\\  \nonumber
    \langle-1|H''_{M}|-1\rangle=\frac{|G_1|^2}{-\gamma_e B+D(\theta)}-\frac{|G_2|^2}{2\gamma_e B}.
\end{eqnarray}
After shifting the zero of energy, the effective Hamiltonian in the laboratory frame is given by
 \begin{eqnarray}
    H''_{eff} = \gamma_{e}BS_{z}+D(\theta)S_{z}^{2}+\delta(\theta)S_z, \nonumber
\end{eqnarray}
in which
$$\delta(\theta)=\frac{\gamma_e B|G_1|^2}{(\gamma_e B)^2-[D(\theta)]^2}+\frac{|G_2|^2}{2\gamma_e B}.$$
Therefore a strong magnetic field suppresses the first order effect of the off-diagonal terms in
$H''_{NV}$ but results in a second order modification of the diagonal elements $\delta(\theta)$.
This is the Hamitonian as in eq. (\ref{hamNV}).

\begin{figure}
\center
\includegraphics[width=2.8in]{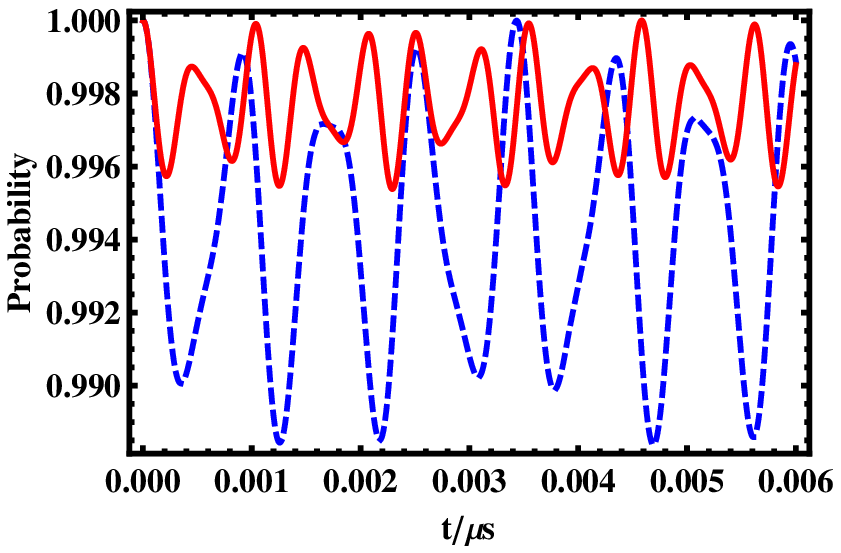}
\caption{Validation of neglecting off-resonant coupling with $D=(2\pi)2870$ MHz, $\theta=10^\circ$, $E=(2\pi)20$ MHz.
The dashed blue line and solid red line present the time evolutions of the probability of the state $|0\rangle$ corresponding
to $B=0.36$ T and $B=0.54$ T, respectively. }
\label{figA2}
\end{figure}

In order to estimate the validity of this approximation in the high magnetic field limit we have
prepared the system in state $|m_s=0\rangle$ and observed the subsequent time evolution of its
population under Hamiltonian $H''_{N}$ in Fig. \ref{figA2} (see caption for simulation parameters).
It is evident, that the population of the initial state $m_s=0$ remains essentially constant which
implies in turn that the effect of the off-diagonal elements in Hamiltonian $H''_{NV}$ are indeed
negligible for the moderate to high magnetic field case that is relevant to our work.


\begin{thebibliography}{10}
\bibitem{spin1}
The components of $\vec{S} = S_x {\vec e}_x + S_y {\vec e}_y + S_z {\vec e}_z$
  are given by \begin{displaymath} S_x =
  \frac{1}{\sqrt{2}}\left(\begin{array}{ccc} 0 & 1 & 0\\ 1 & 0 & 1\\ 0 & 1 &
  0\end{array}\right), S_y = \frac{1}{\sqrt{2}}\left(\begin{array}{ccc} 0 & -i
  & 0\\ i & 0 & -i\\ 0 & i & 0\end{array}\right), S_z =
  \left(\begin{array}{ccc} 1 & 0 & 0\\ 0 & 0 & 0\\ 0 & 0 & -1\end{array}\right)
  \end{displaymath}.
\bibitem{Number}
A spherical shaped nanodiamond of diameter 30 nm has the volume
  $3.21\times10^{-17}$ $cm^3$. Each unit cell of the diamond crystal has the
  volume $4.49\times10^{-23}$ $cm^3$, the number of unit cells in the given
  particle is $7.14\times10^6$. 8 carbon atoms per unit cell of diamond crystal
  gives the total number of carbon atom $5.71\times10^7$. Natural abundance of
  the $^{13}C$ 1.1\% leads to the total number of $^{13}C$ nuclear spin in a 30
  nm nanodiamond is $2.29\times10^5$.
\end{thebibliography}

\end{document}